\documentclass[journal,twocolumn,10pt]{IEEEtran}
\pdfoutput=1

\usepackage{amsmath,amssymb,amsfonts,amsthm}
\usepackage{bm}
\usepackage{graphicx}
\usepackage{booktabs}
\usepackage{multirow}
\usepackage{array}
\usepackage{cite}
\usepackage{url}
\usepackage{xcolor}
\usepackage{algorithm}
\usepackage{algorithmic}

\newtheorem{theorem}{Theorem}
\newtheorem{proposition}{Proposition}

\newtheorem{corollary}{Corollary}

\newcommand{\btheta}{\boldsymbol{\theta}}
\newcommand{\bPhi}{\boldsymbol{\Phi}}
\newcommand{\bh}{\mathbf{h}}
\newcommand{\bs}{\mathbf{s}}
\newcommand{\bx}{\mathbf{x}}
\newcommand{\by}{\mathbf{y}}
\newcommand{\bn}{\mathbf{n}}
\newcommand{\br}{\mathbf{r}}
\newcommand{\bH}{\mathbf{H}}
\newcommand{\bT}{\mathbf{T}}
\newcommand{\bI}{\mathbf{I}}
\newcommand{\bJ}{\mathbf{J}}
\newcommand{\bP}{\mathbf{P}}
\newcommand{\bQ}{\mathbf{Q}}
\newcommand{\bD}{\mathbf{D}}
\newcommand{\bA}{\mathbf{A}}
\newcommand{\bB}{\mathbf{B}}

\newcommand{\cC}{\mathcal{C}}
\newcommand{\cG}{\mathcal{G}}
\newcommand{\cS}{\mathcal{S}}
\newcommand{\cF}{\mathcal{F}}
\newcommand{\CN}{\mathcal{CN}}

\newcommand{\VTwoParams}{8.1K}
\newcommand{\VTwoPureParams}{5.7K}
\newcommand{\VTwoWiSigEqZero}{73.80$\pm$0.77}
\newcommand{\VTwoWiSigEqOne}{55.03$\pm$1.03}
\newcommand{\VTwoWiFiRobAll}{82.18$\pm$0.39}
\newcommand{\VTwoWiFiRobMM}{83.40$\pm$0.20}
\newcommand{\VTwoWiFiRobMS}{81.77$\pm$0.45}
\newcommand{\VTwoWiFiRobFourAnt}{86.39$\pm$0.34}

\newcommand{\VTwoMobCross}{34.01$\pm$1.85}
\newcommand{\VTwoLoRaMultiAll}{82.40$\pm$0.45}
\newcommand{\VTwoLoRaDay}{81.93$\pm$0.59}
\newcommand{\VTwoLoRaLoc}{84.70$\pm$0.62}
\newcommand{\VTwoLoRaRX}{81.99$\pm$0.63}
\newcommand{\VTwoLoRaAug}{84.20$\pm$0.63}
\newcommand{\VTwoLoRaTwelveCh}{92.15$\pm$0.41}
\newcommand{\VTwoLoRaMultiSFSame}{83.85$\pm$1.07}
\newcommand{\VTwoSimR}{0.981 / 0.993}
\newcommand{\VTwoWiFiSimR}{0.981}
\newcommand{\VTwoLoRaSimR}{0.993}

\newcommand{\VTwoRogueAUROC}{69.06}

\newcommand{\TSLParams}{8.1K}
\newcommand{\TSLWiSigEqZero}{72.66$\pm$0.93}
\newcommand{\TSLWiSigEqOne}{54.17$\pm$0.47}
\newcommand{\TSLWiFiRobAll}{82.18$\pm$0.40}
\newcommand{\TSLWiFiRobMM}{83.33$\pm$0.23}
\newcommand{\TSLWiFiRobMS}{81.79$\pm$0.47}
\newcommand{\TSLWiFiRobFourAnt}{86.41$\pm$0.62}

\newcommand{\TSLMobCross}{34.46$\pm$1.21}
\newcommand{\TSLLoRaMultiAll}{81.89$\pm$0.37}
\newcommand{\TSLLoRaDay}{82.64$\pm$0.40}
\newcommand{\TSLLoRaLoc}{83.99$\pm$0.40}
\newcommand{\TSLLoRaRX}{80.80$\pm$0.69}
\newcommand{\TSLLoRaAug}{84.57$\pm$1.15}
\newcommand{\TSLLoRaTwelveCh}{92.14$\pm$1.28}
\newcommand{\TSLLoRaMultiSFSame}{84.38$\pm$1.10}
\newcommand{\TSLSimR}{0.980 / 0.993}

\newcommand{\UTGParams}{1.6K}
\newcommand{\UTGWiSigEqZero}{51.37$\pm$0.13}
\newcommand{\UTGWiSigEqOne}{30.87$\pm$0.15}
\newcommand{\UTGWiFiRobAll}{68.06$\pm$2.25}
\newcommand{\UTGWiFiRobMM}{71.19$\pm$1.95}
\newcommand{\UTGWiFiRobMS}{67.02$\pm$2.35}
\newcommand{\UTGWiFiRobFourAnt}{83.96$\pm$0.40}

\newcommand{\UTGMobCross}{22.65$\pm$0.75}
\newcommand{\UTGLoRaMultiAll}{81.81$\pm$1.38}
\newcommand{\UTGLoRaDay}{85.30$\pm$1.77}
\newcommand{\UTGLoRaLoc}{83.18$\pm$0.67}
\newcommand{\UTGLoRaRX}{79.31$\pm$1.55}
\newcommand{\UTGLoRaAug}{84.62$\pm$0.63}
\newcommand{\UTGLoRaTwelveCh}{91.94$\pm$0.25}
\newcommand{\UTGLoRaMultiSFSame}{84.06$\pm$1.10}
\newcommand{\UTGSimR}{0.970 / 0.959}

\newcommand{\TETParams}{14.4K}
\newcommand{\TETWiSigEqZero}{58.14$\pm$0.61}
\newcommand{\TETWiSigEqOne}{35.95$\pm$1.97}
\newcommand{\TETWiFiRobAll}{69.07$\pm$0.80}
\newcommand{\TETWiFiRobMM}{73.38$\pm$0.64}
\newcommand{\TETWiFiRobMS}{67.64$\pm$0.89}
\newcommand{\TETWiFiRobFourAnt}{85.00$\pm$0.92}

\newcommand{\TETMobCross}{30.15$\pm$0.78}
\newcommand{\TETLoRaMultiAll}{76.28$\pm$2.94}
\newcommand{\TETLoRaDay}{89.52$\pm$3.64}
\newcommand{\TETLoRaLoc}{82.61$\pm$3.13}
\newcommand{\TETLoRaRX}{66.44$\pm$4.78}
\newcommand{\TETLoRaAug}{83.36$\pm$1.33}
\newcommand{\TETLoRaTwelveCh}{90.82$\pm$1.09}
\newcommand{\TETLoRaMultiSFSame}{86.43$\pm$1.89}
\newcommand{\TETSimR}{--- / ---}

\newcommand{\BrZeroWiSig}{74.30$\pm$0.69}
\newcommand{\BrZeroWiFi}{58.76$\pm$0.46}
\newcommand{\BrZeroLoRaLoc}{51.10$\pm$2.52}
\newcommand{\BrZeroLoRaCh}{21.06$\pm$8.34}
\newcommand{\BrOneWiSig}{53.40$\pm$1.12}
\newcommand{\BrOneWiFi}{63.07$\pm$0.22}
\newcommand{\BrOneLoRaLoc}{54.95$\pm$2.07}
\newcommand{\BrOneLoRaCh}{63.06$\pm$3.40}
\newcommand{\BrTwoWiSig}{53.62$\pm$0.64}
\newcommand{\BrTwoWiFi}{67.81$\pm$0.30}
\newcommand{\BrTwoLoRaLoc}{82.22$\pm$0.47}
\newcommand{\BrTwoLoRaCh}{87.14$\pm$0.20}
\newcommand{\BrThreeWiSig}{52.07$\pm$2.40}
\newcommand{\BrThreeWiFi}{72.40$\pm$0.33}
\newcommand{\BrThreeLoRaLoc}{85.09$\pm$0.59}
\newcommand{\BrThreeLoRaCh}{92.62$\pm$0.62}
\newcommand{\BrFourWiSig}{72.33$\pm$1.95}
\newcommand{\BrFourWiFi}{77.34$\pm$0.62}
\newcommand{\BrFourLoRaLoc}{84.70$\pm$0.62}
\newcommand{\BrFourLoRaCh}{92.15$\pm$0.41}

\newcommand{\AblNoOrthWiSig}{46.12$\pm$1.03}
\newcommand{\AblNoOrthWiFi}{65.55$\pm$0.39}
\newcommand{\AblNoOrthLoRaLoc}{75.29$\pm$0.59}
\newcommand{\AblNoOrthLoRaCh}{65.45$\pm$1.06}
\newcommand{\AblNoToepWiSig}{73.20$\pm$0.78}
\newcommand{\AblNoToepWiFi}{72.43$\pm$0.64}
\newcommand{\AblNoToepLoRaLoc}{50.98$\pm$1.83}
\newcommand{\AblNoToepLoRaCh}{78.26$\pm$1.94}
\newcommand{\AblNoGaugeWiSig}{73.40$\pm$0.89}
\newcommand{\AblNoGaugeWiFi}{76.48$\pm$0.37}
\newcommand{\AblNoGaugeLoRaLoc}{84.66$\pm$0.27}
\newcommand{\AblNoGaugeLoRaCh}{92.06$\pm$0.53}
\newcommand{\AblNoChirpWiSig}{73.14$\pm$1.34}
\newcommand{\AblNoChirpWiFi}{77.26$\pm$0.64}
\newcommand{\AblNoChirpLoRaLoc}{79.63$\pm$0.42}
\newcommand{\AblNoChirpLoRaCh}{89.65$\pm$0.69}
\newcommand{\AblNoChipWiSig}{72.07$\pm$2.04}
\newcommand{\AblNoChipWiFi}{76.62$\pm$0.99}
\newcommand{\AblNoChipLoRaLoc}{79.18$\pm$0.62}
\newcommand{\AblNoChipLoRaCh}{83.82$\pm$0.89}

\newcommand{\AblNoAugWiSig}{65.50$\pm$1.76}
\newcommand{\AblNoAugWiFi}{72.31$\pm$0.71}
\newcommand{\AblNoAugLoRa}{93.91$\pm$0.26}
\newcommand{\AblSigMixWiSig}{69.70$\pm$2.15}
\newcommand{\AblSigMixWiFi}{73.19$\pm$0.95}
\newcommand{\AblSigMixLoRa}{93.70$\pm$0.56}
\newcommand{\AblKThreeWiSig}{75.97$\pm$2.85}
\newcommand{\AblKThreeWiFi}{77.09$\pm$1.13}
\newcommand{\AblKThreeLoRa}{79.68$\pm$2.17}
\newcommand{\AblKFiveWiSig}{73.73$\pm$1.50}
\newcommand{\AblKFiveWiFi}{75.03$\pm$1.94}
\newcommand{\AblKFiveLoRa}{78.74$\pm$1.84}

\begin{document}

\title{Gauge-Invariant Bilinear Deep Unfolding for Cross-Domain Specific Emitter Identification: Identifiability, Cram\'er-Rao Bounds, and Linear Convergence}

\author{Fengchong~Yao, Jianbing~Li, Qing~Liu, Kefeng~Song, Haitao~Li, Song~Wang, and Kunpeng~Wang%
\thanks{Fengchong Yao, Jianbing Li, Qing Liu, Kefeng Song, Haitao Li, Song Wang, and Kunpeng Wang are with the School of Information Systems Engineering, Information Engineering University, Zhengzhou 450001, China (e-mail: phoenixly@126.com; li\_jb@126.com; liuqing8123@163.com; annx1990@163.com; lihaitao\_01@163.com; wangsong8190@163.com; wkp0110@163.com).}%
\thanks{Corresponding author: Jianbing Li (e-mail: li\_jb@126.com).}%
\thanks{This work has been submitted to the IEEE for possible publication. Copyright may be transferred without notice, after which this version may no longer be accessible.}}

\markboth{PREPRINT}%
{YAO \MakeLowercase{\textit{et al.}}: GAUGE-INVARIANT BILINEAR DEEP UNFOLDING FOR CROSS-DOMAIN SPECIFIC EMITTER IDENTIFICATION}

\maketitle

\begin{abstract}
Specific emitter identification (SEI) authenticates wireless transmitters via microscopic RF impairments. Despite high in-domain accuracy, black-box networks collapse across days, rooms, and receivers because multipath channels ($\sim 0\text{ dB}$) overwhelm intrinsic nonlinearities ($\sim -30\text{ dB}$). Unlike linear deep unfolding with known operators, cross-domain SEI poses a bilinear inverse problem: jointly separating packet multipath channels and shared wideband RF kernels from preambles under unobservable $U(1)_+ \times U(1)_-$ carrier and mirror phase gauges. We propose \textbf{GaugeInvariantADUNet}, an \VTwoParams{}-parameter three-layer physics-anchored unfolding architecture uniting a \VTwoPureParams{} analytical white-box core with an optional $2.4\text{K}$ channel-orthogonal residual branch. It combines (i) a charge-graded orthogonal dictionary with multiresolution chirp bases unifying Wi-Fi and LoRa waveforms; (ii) an exact Toeplitz Cholesky solver with Parseval-Jacobi preconditioning eliminating $13.8\text{ dB}$ circular-FFT stopband bias; and (iii) a three-block torus gauge-invariant readout with chipset centering preventing conjugate phase flips. Theoretically, we prove subspace identifiability, the Schur-complement Cram\'er-Rao bound, non-alternating LISTA asymptotic leakage bounds, quotient completeness, and local R-linear convergence. Across 100+ Wi-Fi and LoRa transfers, GaugeInvariantADUNet outperforms 13 baselines with $102\times$ fewer parameters than equalized ResNets. Code will be released upon acceptance.
\end{abstract}


\section{Introduction}
\label{sec:intro}

\IEEEPARstart{P}{hysical-layer} device authentication via specific emitter identification (SEI), also known as radio-frequency fingerprinting (RFFI), identifies wireless transmitters directly from analog waveform distortions induced by manufacturing tolerances in digital-to-analog converters (DACs), quadrature I/Q mixers, phase-locked loops (PLLs), and power amplifiers (PAs) \cite{danev2012physical, jagannath2022comprehensive, shen2022towards, brik2008wireless, polak2011identifying}. Classical RF behavioral modeling characterizes these front-end imperfections via Volterra series, generalized memory polynomials (GMP), and frequency-selective I/Q imbalance models \cite{ding2004robust, morgan2006generalized, valkama2001advanced, peng2019design}. Because silicon-level imperfections cannot be forged via Medium Access Control (MAC) spoofing, SEI offers zero-signaling-overhead authentication for Internet-of-Things (IoT) networks (e.g., LoRaWAN) and wireless local-area networks (IEEE 802.11 Wi-Fi) \cite{hanna2022wisig, shen2021radio, riyaz2018deep}.

However, deploying deep-learning-based SEI in dynamic propagation environments faces a notorious obstacle: \emph{cross-domain performance collapse} \cite{al2020exposing, hanna2022wisig, shen2023towards, restuccia2019deepradioid, sankhe2019oracle}. Trained on single-day static captures, end-to-end convolutional neural networks (CNNs) \cite{merchant2018deep, soltani2020more}, complex-valued neural networks (CVNNs) \cite{trabelsi2018deep}, Patch Transformers \cite{dosovitskiy2021image}, and Kolmogorov--Arnold networks (KANs) \cite{liu2025kan} latch onto two high-energy shortcuts: (i) temperature-drifting carrier frequency offset (CFO), and (ii) the quasi-static multipath impulse response $\bh_k$ linking a benchtop transmitter to the receiver. Because linear multipath alters waveform amplitude and phase at $O(1)$ relative energy ($0\text{ dB}$), whereas genuine emitter nonlinearities reside $-25\text{ dB}$ to $-40\text{ dB}$ below the linear carrier, black-box networks memorize channel transfer functions and suffer $30\%\text{--}70\%$ accuracy drops across days, rooms, walking pedestrians, or heterogeneous software-defined radios (SDRs) \cite{mazokha2025mobrffi, cekic2021robust, zhang2023radio}.

\subsection{Related Work and Fundamental Limitations}
Existing approaches to channel-robust SEI and structured inverse recovery fall into four paradigms:

\subsubsection{Data Augmentation and Heuristic Spectral Representations}
Virtual multipath augmentation \cite{soltani2020more, cekic2021robust}, channel-independent spectrograms (CIS) \cite{shen2022towards}, inactive-subcarrier and out-of-band spectral regrowth CNNs (SR-CNN) \cite{xie2024towards}, bispectral CNNs \cite{ding2018specific}, and cross-domain amplitude spectrum mixing (SigMix) \cite{wan2025sigmix} regularize black-box backbones invariance-wise. Yet under strict zero-augmentation (\texttt{no\_aug}) source training or non-line-of-sight (NLOS) through-wall propagation, over-parameterized backbones ($90\text{K}\text{--}830\text{K}$ parameters) still entangle residual multipath dispersion with emitter features while offering zero parameter-level physical verifiability.

\subsubsection{Two-Stage Equalization with Black-Box CNNs or Linear LISTA}
Given known protocol preambles $\bs[n]$ (IEEE 802.11 L-STF/L-LTF or Semtech LoRa upchirps), two-stage pipelines \cite{peng2019design, zhang2023radio} estimate a single-pass linear channel $\hat{\bh}_{\mathrm{LS}}$ solely from $\bs[n]$, equalize the received packet, and feed the residual into a ResNet or linear sparse coder. However, single-pass channel estimation treats emitter nonlinearity $\bPhi(\bs)\btheta$ as colored noise, incurring an irreducible second-order cross-leakage bias $(\Delta\bh)*(\bPhi\btheta)$ under frequency-selective multipath (formalized in Corollary~\ref{cor:twostage}), and suffers a $13.8\text{ dB}$ stopband-nulling bias when circular FFT division is applied to oversampled narrowband waveforms ($8\times$-oversampled LoRa CSS) or guard-band OFDM preambles.

\subsubsection{Bilinear Inverse Problems and Model-Driven Deep Unfolding}
Blind and semi-blind deconvolution has been extensively studied in signal processing via convex lifting \cite{ahmed2014blind}, subspace-constrained sparse recovery \cite{ling2015self, ling2017blind}, and non-convex alternating gradient descent with local linear convergence \cite{li2019rapid, chi2019nonconvex}. In parallel, following learned ISTA (LISTA) \cite{gregor2010learning} and its theoretical linear-convergence analyses \cite{chen2018theoretical, liu2019alista}, algorithm unfolding has excelled in MIMO detection (DetNet \cite{samuel2019learning}, OAMP-Net \cite{he2020model}), learned AMP \cite{borgerding2017amp}, ADMM-Net \cite{yang2016deep}, and compressive channel estimation \cite{balatsoukas2019deep, monga2021algorithm, shlezinger2023model}. \emph{Crucially, standard wireless unfolding solves linear inverse problems $\by = \bA\btheta + \bn$ with a known sensing matrix $\bA$.} In contrast, cross-domain SEI requires solving a \textbf{bilinear semi-blind inverse problem} $\tilde{\by}_k = \bh_k * (\bs + \bPhi(\bs)\btheta) + \bn_k$, where \emph{both} the $L_h$-tap multipath channel $\bh_k$ and the $M$-dimensional nonlinear RF kernel $\btheta$ are unknown and multiplicatively coupled across protocols with incompatible excitation physics---high peak-to-average power ratio (PAPR) in Wi-Fi OFDM versus unit constant envelope ($|s[n]|\equiv 1$) in LoRa CSS, on which standard PA polynomials $|s[n]|^{2p}s[n] \equiv s[n]$ degenerate into linear constants inside the multipath subspace.

\subsubsection{Group-Invariant Signal Representations}
Group-invariant descriptors---such as wavelet scattering transforms \cite{mallat2012group, bruna2013invariant}, bispectral invariants on compact groups \cite{kakarala2012bispectrum}, and orbit-recovery polynomial invariants \cite{bandeira2017estimation, bendory2023compact}---characterize signals modulo symmetry group actions. Conventional complex unfolding readouts concatenate raw Cartesian coordinates $[\Re(\btheta), \Im(\btheta)]$, violating \textbf{$U(1)_+ \times U(1)_-$ carrier-mirror gauge invariance}: whenever propagation geometry, packet trigger slicing, or quadrature conversion introduces an unobservable phase offset $e^{j\phi_0}$ between the co-rotating signal path ($\propto \bs$, charge $+1$) and the conjugate-mirror path ($\propto \bs^*$, charge $-1$), recovered mirror modes rotate by $e^{-j2\phi_0}$ and flip feature signs across rooms and SDRs.

\subsection{Summary of Contributions}
We propose \textbf{GaugeInvariantADUNet} (\VTwoPureParams{} parameters in pure analytical white-box mode; \VTwoParams{} parameters with the optional $2.4\text{K}$ bounded orthogonal residual compensator at $T=3$), making four main contributions:
\begin{enumerate}
    \item \textbf{Charge-Graded Multi-Resolution Bilinear Unfolding via Exact Two-Sided Toeplitz Solver}: We formulate preamble-assisted SEI on the orthogonal complement $\cC^\perp$ of the two-sided multipath subspace. We construct a charge-graded dictionary $\bPhi(\bs)=[\bPhi_+(\bs), \bPhi_-(\bs)]$ unifying OFDM PA memory with multi-resolution ($P/2, P/4, P/8$) LoRa chirp differential bases and sweep harmonics, solved via $\mathcal{O}(L\log L + L_h^2)$ two-sided Toeplitz Cholesky updates and Parseval Jacobi preconditioning that converge within $T=3$ layers.
    \item \textbf{Minimal Quotient-Complete $U(1)_+ \times U(1)_-$ Gauge Readout \& Bounded Orthogonal Residual Compensator}: Pairing charge-preserving diagonally dominant block preconditioners $\operatorname{blkdiag}(\bP_+^{(t)}, \bP_-^{(t)})$ with a minimal 3-block 1-homogeneous readout $\cG(\btheta)$ (moduli $|\btheta|$, carrier-anchored proper modes $Z_+$, and all-pairs mirror cross-spectra $H_-$) plus chipset common-mode centering eliminates cross-domain phase flips with exact quotient completeness. For high-PAPR OFDM waveforms exhibiting non-smooth PA/DAC clipping beyond finite-degree dictionaries, we derive an explicit $2.4\text{K}$-parameter $\bP_{\cC}^\perp$-constrained, $U(1)$-invariant residual compensator $\mathbf{z}_{\mathrm{res}}$ while preserving purely analytical waveform synthesis.
    \item \textbf{Identifiability, Schur-Complement CRLB, Two-Stage Leakage Floor, and Linear Convergence}:
    \begin{itemize}
        \item \emph{Theorem 1 \& Corollary 1}: Using Wirtinger complex calculus \cite{schreier2010statistical, vandenbos1994complex}, we derive the holomorphic FIM and exact Schur-complement nuisance-projected CRLB $\operatorname{Tr}(\bJ_{\mathrm{eff}}(\btheta)^{-1})$, prove necessary and sufficient subspace disjointness conditions, and prove why non-alternating two-stage LISTA hits an asymptotic high-SNR cross-leakage floor.
        \item \emph{Theorem 2}: We prove exact invariance under complex channel scaling/rotation and arbitrary torus rotations $(\psi_+, \psi_-) \in U(1)_+ \times U(1)_-$, alongside minimal quotient orbit completeness.
        \item \emph{Theorem 3}: We prove local R-linear error contraction in the Euclidean norm under diagonally dominant block preconditioning, governed by the spectral conditioning of $\bJ_{\mathrm{eff}}(\btheta^\star)$.
    \end{itemize}
    \item \textbf{Comprehensive 5-Seed Validation, Constructive Attribution \& 13 Counterpart Baselines}: Across $>100$ zero-shot cross-domain Wi-Fi and LoRa tasks, two controlled internal variants (\texttt{Ours-Unlearned}, \texttt{Ours-Frozen-$\hat{\bh}_{\mathrm{LS}}$}), two equalized/circular unfolding counterparts (\texttt{ToeplitzEq-TinyResNet1D}, \texttt{Circular-FFT ADU-Net}), and nine classical, spectral, and deep baselines, GaugeInvariantADUNet establishes a new accuracy-complexity Pareto frontier at \VTwoParams{} parameters ($102\times$ smaller and $+12.7\%$ F1 higher than the strongest external prior baseline) while providing direct circuit-level physical interpretability ($R^2 = \VTwoWiFiSimR / \VTwoLoRaSimR$).
\end{enumerate}

\section{Bilinear Physical Forward Model and Symmetry Structure}
\label{sec:model}

\subsection{Discrete Baseband Preamble Observation Equation}
Consider an emitter transmitting frames whose preamble corresponds to a known unit-RMS discrete complex baseband sequence $\bs \in \mathbb{C}^L$ ($\|\bs\|_2^2 = L$). Given $K \ge 1$ frames ($K=1$ for standard single-packet SEI; $K \in \{3, 5\}$ for multi-packet collaborative inference), the $k$-th received frame $\by_k \in \mathbb{C}^L$ is
\begin{equation}
    \by_k = \bD(\omega_k)\Big(\bh_k * \cF_{\mathrm{TX}}\big(\bs;\, \btheta\big)\Big) + \bn_k, \quad k = 1, \dots, K,
    \label{eq:raw_obs}
\end{equation}
where $\bD(\omega_k) = \operatorname{diag}(1, e^{j\omega_k}, \dots, e^{j\omega_k(L-1)})$ denotes temperature-drifting LO CFO, $\bh_k \in \mathbb{C}^{L_h}$ ($L_h = L_{\mathrm{pre}} + L_{\mathrm{post}} \ll L$) is the two-sided finite-support multipath channel over lags $[-L_{\mathrm{pre}}, L_{\mathrm{post}})$, and $\bn_k \sim \CN(\mathbf{0}, \sigma^2 \bI_L)$.

\begin{proposition}[Lossless Periodic Preamble CFO Decoupling]
\label{prop:cfo}
If $s[n + P] = s[n]$ over $n \in [n_0, N_{\mathrm{per}} - P)$ (e.g., $P = 16$ or $20$ on Wi-Fi L-STF; $P = 2^{\mathrm{SF}} f_s / B_{\mathrm{chirp}}$ on LoRa upchirps), any time-invariant memory-bounded nonlinear TX operator $\cF_{\mathrm{TX}}$ cascaded with quasi-static $\bh_k$ preserves period $P$ for $n \ge n_0$. Thus $\hat{\omega}_k = \frac{1}{P}\angle \sum_{n=n_0}^{N_{\mathrm{per}}-P-1} y_k^*[n] y_k[n+P]$ strips $\bD(\omega_k)$ via $\tilde{\by}_k = \bD(-\hat{\omega}_k)\by_k$ without distorting emitter harmonics.
\end{proposition}
\begin{proof}
By time-invariance and finite memory $M_{\mathrm{mem}}$ of $\cF_{\mathrm{TX}}$, $s[n+P]=s[n]$ implies $x_{\mathrm{TX}}[n+P]=x_{\mathrm{TX}}[n]$ over the steady-state window, and linear convolution with quasi-static $\bh_k$ preserves $r_k^{\mathrm{ch}}[n+P]=r_k^{\mathrm{ch}}[n]$. Under $y_k[n]=e^{j\omega_k n}r_k^{\mathrm{ch}}[n]$ (noiseless limit), $\sum_{n=n_0}^{N_{\mathrm{per}}-P-1} y_k^*[n]y_k[n+P] = e^{j\omega_k P}\sum_n |r_k^{\mathrm{ch}}[n]|^2$, recovering $\hat{\omega}_k=\omega_k$ whenever $|\omega_k P|<\pi$ without distorting $\bh_k$ or $\bPhi(\bs)\btheta$.
\end{proof}

After unit-RMS normalization and CFO derotation, expanding the RF front-end distortion over an $M$-atom physical dictionary $\bPhi(\bs) \in \mathbb{C}^{L \times M}$ (plus optional unmodeled orthogonal parasitic ripple $\mathbf{e}_\perp \in \cC^\perp$) yields the bilinear semi-blind model:
\begin{equation}
    \tilde{\by}_k = \bH_k\big(\bs + \bPhi(\bs)\btheta\big) + \bn_k = \bT(\bx_{\btheta})\bh_k + \bn_k,
    \label{eq:bilinear_obs}
\end{equation}
where $\bx_{\btheta} \triangleq \bs + \bPhi(\bs)\btheta$ and $[\bT(\bx)]_{n, l} = x[n + L_{\mathrm{pre}} - l]$.

\subsection{Charge-Graded Multi-Resolution Physical Dictionary}
\label{sec:basis}

Wi-Fi OFDM exhibits $>10\text{ dB}$ envelope fluctuations, whereas LoRa CSS has constant unit envelope ($|s[n]| \equiv 1$), causing conventional PA polynomials $s[n-m]|s[n-m]|^{2p}$ to collapse into shifts $s[n-m] \in \operatorname{Range}(\bT(\bs))$ indistinguishable from multipath. Moreover, under an unobservable trigger/carrier phase shift $\bs \mapsto e^{j\phi_0}\bs$, physical nonlinearities split into two irreducible $U(1)$ representations of \textbf{phase charge} $q \in \{+1, -1\}$:
\begin{equation}
    \bPhi_\pm(e^{j\phi_0}\bs) = e^{\pm j\phi_0}\bPhi_\pm(\bs).
    \label{eq:charge_def}
\end{equation}

Let $\cC \triangleq \operatorname{Range}(\bT(\bs))$ have orthonormal basis $\bQ_{\mathrm{ch}} \in \mathbb{C}^{L \times L_h}$ and orthogonal complement projector $\bP_{\cC}^\perp \equiv \bP_{\cC^\perp} = \bI_L - \bQ_{\mathrm{ch}}\bQ_{\mathrm{ch}}^H$. We construct $\bPhi(\bs) = [\bPhi_+(\bs), \, \bPhi_-(\bs)] \in \mathbb{C}^{L \times (M_+ + M_-)}$ via two-pass modified Gram--Schmidt orthonormalization strictly inside $\bP_{\cC}^\perp$ within each charge sector separately ($\bP_{\cC}^\perp \bPhi_\pm = \bPhi_\pm$, $\frac{1}{L}\bPhi_\pm^H \bPhi_\pm = \bI_{M_\pm}$):
\begin{enumerate}
    \item \textbf{Charge $+1$ Proper Nonlinear \& Chirp-Sweep Sector $\bPhi_+(\bs) \in \mathbb{C}^{L \times M_+}$}: (i) Chebyshev PA envelope memory $s[n-m] T_d(\widetilde{|s|}[n-m])$ and bias memory $s[n]|s[n-m]|^2$ (active on variable-envelope OFDM); (ii) PLL instantaneous-frequency Chebyshev and Fourier sine harmonics $s[n] T_d(f_{\mathrm{inst}}[n])$, $s[n]\sin(k\pi f_{\mathrm{inst}}[n])$; (iii) \emph{multi-resolution chirp differential bases} $s[n]\Re(g_\delta[n]), s[n]\Im(g_\delta[n])$ built from charge-$0$ chordal products $g_\delta[n] = s[n]s^*[n-\delta]$ at fractional periods $\delta \in \{2, 6, P/16, P/8, P/4, P/2\}$ (illustrated in Fig.~\ref{fig:toeplitz_chirp}(c)); and (iv) chirp wrap-around ringing $s[n]|\Delta^2\phi[n-d]|$ and cubic slew-rate terms $|\Delta s|^2\Delta s$---while excluding non-periodic slow-time slopes $\tilde{t}[n]$ and AGC burst envelopes that leak receiver clock/gain artifacts across heterogeneous SDRs.
    \item \textbf{Charge $-1$ Conjugate-Mirror Sector $\bPhi_-(\bs) \in \mathbb{C}^{L \times M_-}$}: Wideband I/Q modulator imbalance and frequency-selective quadrature filter skew atoms $s^*[n-m]$, $s^*[n]|s[n]|^2$, $s^*[n] f_{\mathrm{inst}}^d[n]$ ($d=1,2$), and $s^*[n]\Im(g_{P/4}[n])$.
\end{enumerate}

\section{GaugeInvariantADUNet Architecture}
\label{sec:arch}

\begin{figure*}[!t]
    \centering
    \includegraphics[width=0.92\textwidth]{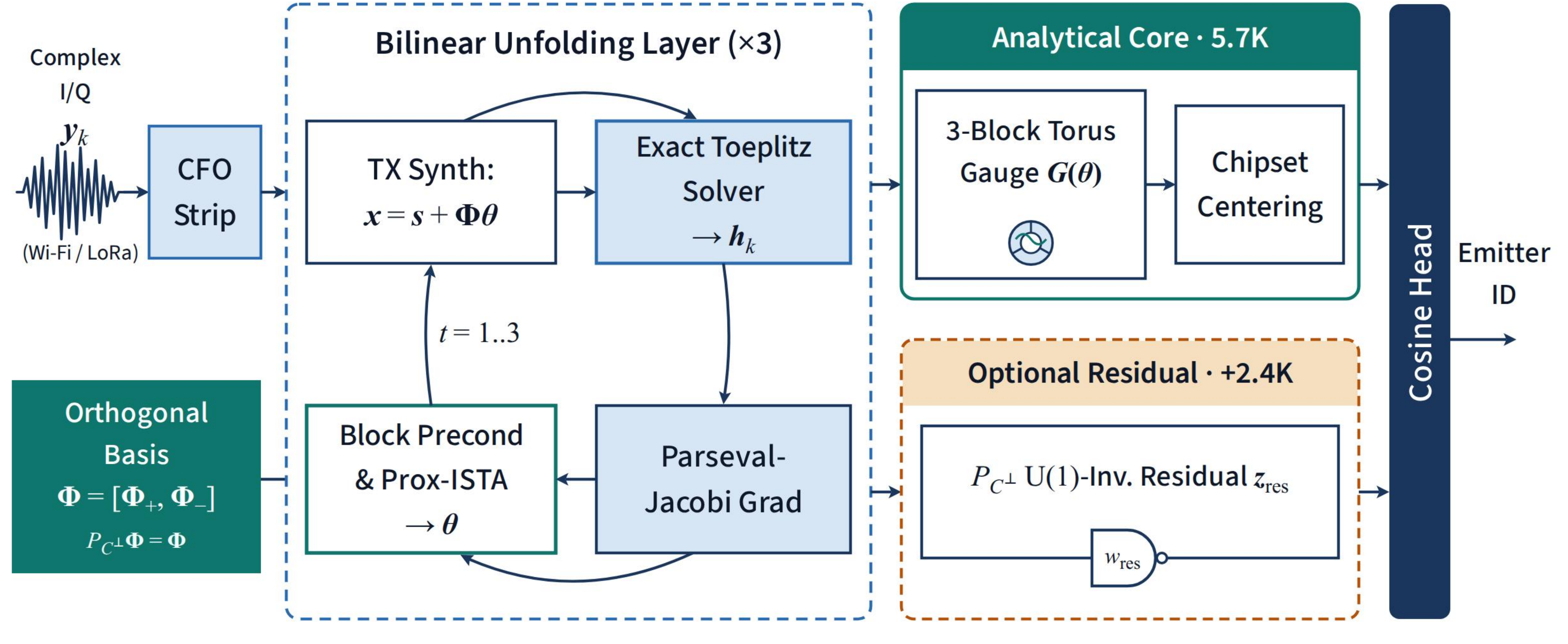}
    \caption{Complete architecture and tensor flow of \texttt{GaugeInvariantADUNet} ($T=3$). Each unfolded bilinear layer $t \in \{1,\dots,T\}$ alternates an exact two-sided Toeplitz Cholesky channel update $\bh_k^{(t)}$ with a Parseval-Jacobi preconditioned, charge-preserving proximal ISTA update on $\btheta^{(t)}$. Downstream, the \VTwoPureParams{}-parameter \emph{Pure Analytical Core} extracts the minimal 3-block $U(1)_+ \times U(1)_-$ gauge-invariant readout $\cG(\btheta^{(T)})$ with chipset common-mode centering, while an optional $2.4\text{K}$-parameter $\bP_{\cC}^\perp$-constrained, $U(1)$-invariant residual branch $\mathbf{z}_{\mathrm{res}}$ captures unmodeled high-PAPR PA/DAC saturation on Wi-Fi without altering purely analytical waveform synthesis $\hat{\by}_k$.}
    \label{fig:arch}
\end{figure*}

We cast joint multi-packet channel deconvolution and sparse RF fingerprint recovery as the bilinear variational problem:
\begin{equation}
\begin{aligned}
    \min_{\{\bh_k \in \mathbb{C}^{L_h}\},\, \btheta \in \mathbb{C}^M} \; &\frac{1}{2K}\sum_{k=1}^K \big\|\tilde{\by}_k - \bT(\bs + \bPhi\btheta)\bh_k\big\|_2^2 \\
    &+ \frac{\mu L}{2}\sum_{k=1}^K \|\bh_k\|_2^2 + \sum_{m=1}^M \tau_m |\theta_m|.
\end{aligned}
    \label{eq:variational}
\end{equation}
Unfolding $T=3$ preconditioned alternating steps yields \textbf{GaugeInvariantADUNet}, whose complete block diagram and forward pass are detailed in Fig.~\ref{fig:arch} and Algorithm~\ref{alg:gadunet}.

\begin{algorithm}[t]
\caption{Forward Pass of \texttt{GaugeInvariantADUNet} ($T=3$)}
\label{alg:gadunet}
\begin{algorithmic}[1]
\REQUIRE $K$-packet I/Q input $\{\by_k\}_{k=1}^K$, known preamble $\bs \in \mathbb{C}^L$, charge-graded basis $\bPhi = [\bPhi_+, \bPhi_-] \in \mathbb{C}^{L \times M}$, $\bQ_{\mathrm{ch}} \in \mathbb{C}^{L \times L_h}$.
\STATE Normalize unit RMS and apply periodic CFO derotation (Prop.~\ref{prop:cfo}) $\to \{\tilde{\by}_k\}_{k=1}^K$.
\STATE Initialize $\btheta^{(0)} \leftarrow \mathbf{0}_M$; precompute $\mathbf{Y}_{\mathrm{pad}} = \cF(\operatorname{Pad}_{L_{\mathrm{pre}}}(\tilde{\by}))$.
\FOR{$t = 1, \dots, T$}
    \STATE \textbf{Waveform synthesis}: $\bx^{(t-1)} \leftarrow \bs + \bPhi\btheta^{(t-1)}$.
    \STATE \textbf{Exact two-sided Toeplitz channel step}: Solve $\big(\bT(\bx^{(t-1)})^H\bT(\bx^{(t-1)}) + \mu^{(t)}L\bI_{L_h}\big)\bh_k^{(t)} = \bT(\bx^{(t-1)})^H\tilde{\by}_k$ via zero-sync FFT autocorrelation + boundary Cholesky \eqref{eq:toeplitz_step}.
    \STATE \textbf{Parseval Jacobi-preconditioned gradient}: Compute $\br_k^{(t)} \leftarrow \tilde{\by}_k - \bT(\bx^{(t-1)})\bh_k^{(t)}$, $D_{k,m}^{(t)} = \|\bH_k^{(t)}\boldsymbol{\phi}_m\|_2^2$ via \eqref{eq:parseval_jacobi}, and $\mathbf{g}^{(t)} \leftarrow \frac{1}{K}\sum_{k=1}^K (\bD_k^{(t)} + \gamma_{\mathrm{jac}}^{(t)}L\bI)^{-1}\bPhi^H (\bH_k^{(t)})^H \br_k^{(t)}$.
    \STATE \textbf{Charge-preserving proximal ISTA}: $\btheta^{(t)} \leftarrow \cS_{\boldsymbol{\tau}^{(t)}}\!\big(\btheta^{(t-1)} + \eta^{(t)}\operatorname{blkdiag}(\bP_+^{(t)}, \bP_-^{(t)})\mathbf{g}^{(t)}\big)$.
\ENDFOR
\STATE Refine $\bh_k^{(T+1)}$ on $\bx_{\mathrm{phys}} = \bs + \bPhi\btheta^{(T)}$, synthesize pure analytical waveform $\hat{\by}_k = \bT(\bx_{\mathrm{phys}})\bh_k^{(T+1)}$, and compute carrier anchor $\theta_{+,0} = \frac{1}{L}\langle \hat{\bx}_{\mathrm{eq}}, \bs\rangle$.
\STATE Extract minimal 3-block gauge-invariant observables $\mathbf{f}_{\mathrm{phys}} = \cG(\btheta_+^{(T)}, \btheta_-^{(T)}, \theta_{+,0})$ via \eqref{eq:gauge_readout}, apply chipset common-mode centering, optionally append bounded $U(1)$-invariant residual code $\mathbf{z}_{\mathrm{res}}$ via \eqref{eq:zres_def}, and output cosine logits.
\end{algorithmic}
\end{algorithm}

\subsection{Unbiased Two-Sided Toeplitz Step \& Parseval Jacobi Preconditioning}

\begin{figure*}[!t]
    \centering
    \includegraphics[width=0.90\textwidth]{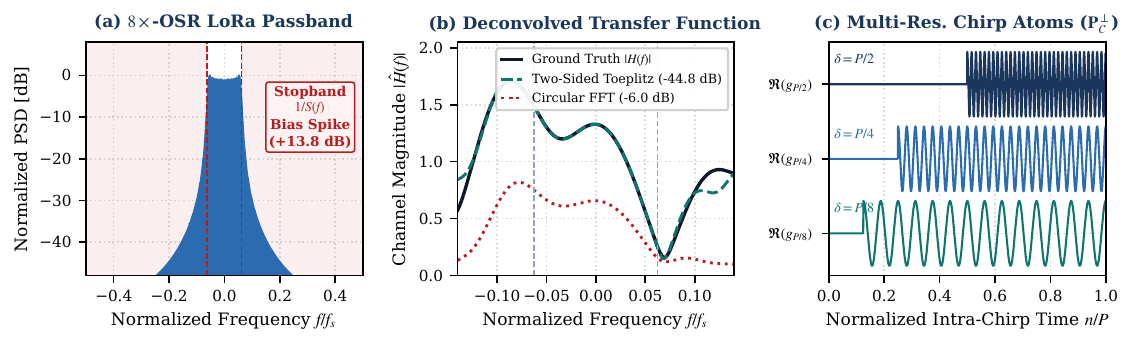}
    \caption{Numerical DSP demonstration of unbiased two-sided Toeplitz deconvolution and multi-resolution orthogonal chirp decomposition on an $8\times$-oversampled LoRa SF7 preamble ($P=1024, L=2048, \mathrm{SNR}=25\text{ dB}$): (a) normalized power spectral density showing the narrow occupied passband ($\eta_{\mathrm{BW}}=1/8$) surrounded by the stopband where circular FFT division $1/S(f)$ incurs $+13.8\text{ dB}$ noise/bias amplification; (b) recovered passband channel transfer magnitude $|\hat{H}(f)|$ comparing exact finite-window two-sided Toeplitz Cholesky \eqref{eq:toeplitz_step} ($-44.8\text{ dB}$ NMSE, tracking ground truth) against biased circular FFT ridge division ($-6.0\text{ dB}$ NMSE); and (c) multi-resolution fractional-period chirp differential bases $\Re(g_\delta[n])$ at $\delta \in \{P/2, P/4, P/8\}$ projected onto $\bP_{\cC}^\perp$.}
    \label{fig:toeplitz_chirp}
\end{figure*}

As demonstrated numerically in Fig.~\ref{fig:toeplitz_chirp}(a)--(b), conventional frequency-domain deconvolution (including \texttt{Circular-FFT ADU-Net} and \texttt{LSEqualized-ResNet1D}) estimates the channel via circular FFT division $\Pi_{[0, L_h)}\cF^{-1}\big\{\frac{\cF(\bx)^* \odot \cF(\tilde{\by})}{|\cF(\bx)|^2 + \mu}\big\}$. Whenever $\bx$ occupies a fraction $\eta_{\mathrm{BW}} = B_{\mathrm{sig}}/f_s < 1$ of the Nyquist band ($\eta_{\mathrm{BW}} = 1/8$ on $1\text{ MS/s}$ LoRa; $\eta_{\mathrm{BW}} = 52/80$ on $25\text{ MS/s}$ Wi-Fi), circular FFT division amplifies thermal noise across the $(1 - \eta_{\mathrm{BW}})$ stopband while imposing periodic head-to-tail boundary wrap-around, convolving the true impulse response with a wide $\operatorname{sinc}$ kernel and stalling waveform reconstruction NMSE at $-7.6\text{ to }-10.7\text{ dB}$.

In Layer $t$ of GaugeInvariantADUNet (Fig.~\ref{fig:arch}), we instead solve the exact finite-support two-sided Toeplitz normal equations:
\begin{equation}
    \bh_k^{(t)} = \Big(\bT(\bx^{(t-1)})^H \bT(\bx^{(t-1)}) + \mu^{(t)} L \bI_{L_h}\Big)^{-1} \bT(\bx^{(t-1)})^H \tilde{\by}_k,
    \label{eq:toeplitz_step}
\end{equation}
in $\mathcal{O}(L\log L + L_h^2)$ operations by extracting lags of $\cF^{-1}\{|\cF(\bx^{(t-1)})|^2\}$, subtracting exact $L_{\mathrm{pre}}$-left and $(L_{\mathrm{post}}-1)$-right boundary rank-update matrices, and solving via batched Cholesky factorization ($L_h=32, L_{\mathrm{pre}}=8$ on full preambles). Next, we apply Parseval's identity to evaluate exact frequency-selective atom curvature in $\mathcal{O}(M)$ operations from cached $|\hat{H}_k^{(t)}[\nu]|^2$ and precomputed $|\hat{\Phi}_m[\nu]|^2$:
\begin{equation}
    D_{k,m}^{(t)} = \frac{1}{N_{\mathrm{FFT}}}\sum_{\nu=0}^{N_{\mathrm{FFT}}-1} \big|\hat{H}_k^{(t)}[\nu]\big|^2 \big|\hat{\Phi}_m[\nu]|^2,
    \label{eq:parseval_jacobi}
\end{equation}
followed by a \emph{diagonally dominant, charge-sector block-diagonal} preconditioner $\bP_{\mathrm{pre}}^{(t)} = \operatorname{blkdiag}(\bP_+^{(t)}, \bP_-^{(t)})$, parameterized explicitly as $\bP_\pm^{(t)} = \operatorname{diag}(\mathbf{d}_\pm^{(t)}) + \delta_{\mathrm{off}}\mathbf{E}_\pm^{(t)}$ with $d_{\pm,m}^{(t)} = 1 + \delta_{\mathrm{diag}}\tanh(a_{\pm,m}^{(t)})$ ($\delta_{\mathrm{diag}} = 0.4$, $\delta_{\mathrm{off}} = 0.12$) and zero-initialized, zero-diagonal bounded complex entries $E_{\pm,m,m'}^{(t)} = \tanh(u_{\pm,m,m'}^{(t)}) + j\tanh(v_{\pm,m,m'}^{(t)})$ ($m\neq m'$). Under zero initialization and weight-decay regularization, off-diagonal coupling remains strictly small ($\|\delta_{\mathrm{off}}\mathbf{E}_\pm^{(t)}\|_2 \le \tilde{\delta}_{\mathrm{off}}$), bounding the spectral deviation by $\|\bP_{\mathrm{pre}}^{(t)} - \bI_M\|_2 \le \delta_{\mathrm{diag}} + \tilde{\delta}_{\mathrm{off}} \triangleq \delta_P$ (used directly in Theorem~\ref{thm:conv}) while preserving exact charge-sector decoupling before element-wise complex soft thresholding $\cS_{\boldsymbol{\tau}^{(t)}}(z) = z \max(0, 1 - \tau^{(t)}/|z|)$.

\emph{Degrees-of-Freedom (DoF) Adaptive Regularization}: Let $\nu_{\mathrm{eff}} = L - L_h$ denote the effective residual degrees of freedom available for joint deconvolution. In the high-DoF full-preamble regime ($\nu_{\mathrm{eff}} \ge 288$, i.e., $L \ge 320$), fixed learnable thresholds $\boldsymbol{\tau}^{(t)}$ prevent over-shrinking attenuated modes in through-wall NLOS propagation. Conversely, in the low-DoF short-window regime ($\nu_{\mathrm{eff}} < 256$, e.g., single-symbol $L=256$ pre-equalized OFDM windows where $L_h$ consumes a significant fraction of observation samples), the solver automatically activates a compact causal channel support ($L_{\mathrm{pre}}=0, L_h \le 16$), scale-free peak-relative Wiener tail shrinkage $h_\ell \leftarrow h_\ell |h_\ell|^2 / (|h_\ell|^2 + \tau_{\mathrm{rel}}^2 \max_{j<2}|h_j|^2)$ ($\tau_{\mathrm{rel}}=0.15$), and orthogonal-residual noise-adaptive thresholding $\boldsymbol{\tau}^{(t)} + \boldsymbol{\gamma}^{(t)}\hat{\sigma}_k$ where $\hat{\sigma}_k = \|\bP_{\cC}^\perp \br_k^{(t)}\|_2 / \sqrt{\nu_{\mathrm{eff}}}$. Crucially, across all regimes, waveform synthesis $\hat{\by}_k = \bT(\bs + \bPhi\btheta^{(T)})\bh_k^{(T+1)}$ relies purely on the analytical physical forward model without black-box decoder MLPs, reaching $-20.9\text{ to }-27.0\text{ dB}$ over-the-air reconstruction NMSE by $T=3$.

\subsection{Minimal Quotient-Complete $U(1)_+ \times U(1)_-$ Gauge Readout \& Bounded Orthogonal Residual Compensator}
\label{sec:readout}

\begin{figure*}[!t]
    \centering
    \includegraphics[width=0.90\textwidth]{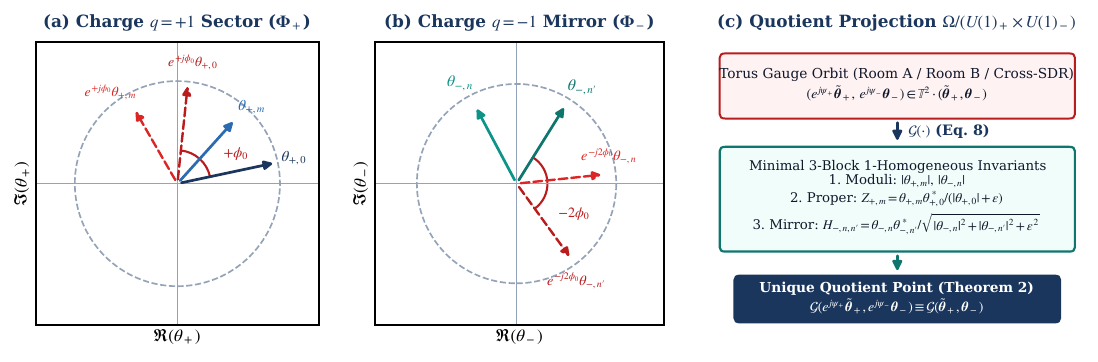}
    \caption{Geometric mechanism of cross-room conjugate phase-flip collapse and minimal torus quotient projection (Theorem~\ref{thm:gauge}): (a) under an unobservable carrier/trigger phase offset $e^{j\phi_0}$, charge $q=+1$ proper modes rotate by $+\phi_0$; (b) simultaneously, charge $q=-1$ conjugate-mirror modes counter-rotate by $-2\phi_0$ relative to the carrier, scrambling raw Cartesian $[\Re,\Im]$ features across rooms and SDRs; and (c) factoring the two-torus $\mathbb{T}^2 = U(1)_+ \times U(1)_-$ action through our minimal 3-block 1-homogeneous readout $\cG(\btheta) = \{|\btheta|, Z_+, H_-\}$ in \eqref{eq:gauge_readout} maps the entire gauge orbit onto a unique invariant quotient point.}
    \label{fig:torus_gauge}
\end{figure*}

As illustrated geometrically in Fig.~\ref{fig:torus_gauge}(a)--(b), under a cross-room propagation or quadrature mixer phase offset $\bs_{\mathrm{eff}} = e^{j\phi_0}\bs$, factoring $e^{j\phi_0}$ into $\bh_k' = e^{j\phi_0}\bh_k$ yields
\begin{equation}
    \cF_{\mathrm{TX}}(e^{j\phi_0}\bs) = e^{j\phi_0}\Big(\bs + \bPhi_+(\bs)\btheta_+ + \bPhi_-(\bs)\big(e^{-j2\phi_0}\btheta_-\big)\Big).
    \label{eq:phase_flip}
\end{equation}
Because the $q=+1$ and $q=-1$ sectors counter-rotate on the complex plane, directly concatenating Cartesian coordinates $[\Re(\btheta), \Im(\btheta)]$ causes cross-room features to jump across decision boundaries. Combined with residual anchor phase $\delta\psi \in [-\pi, \pi)$, the sectors undergo independent rotations $(\tilde{\btheta}_+, \btheta_-) \mapsto (e^{j\psi_+}\tilde{\btheta}_+, e^{j\psi_-}\btheta_-)$ on the two-torus $\mathbb{T}^2 = U(1)_+ \times U(1)_-$ (Fig.~\ref{fig:torus_gauge}(c)), where $\tilde{\btheta}_+ = [\theta_{+,0}, (\btheta_+^{(T)})^T]^T$ includes the carrier anchor $\theta_{+,0} = \frac{1}{L}\langle \hat{\bx}_{\mathrm{eq}}, \bs\rangle$.

We construct the minimal sufficient 1-homogeneous ($\mathcal{O}(|\btheta|)$) gauge-invariant readout $\cG(\tilde{\btheta}_+, \btheta_-) \in \mathbb{R}^{M + 2M_+ + 2\binom{M_-}{2}}$ from \emph{only three} algebraically non-redundant observable blocks ($\epsilon = 10^{-6}$):
\begin{align}
    \text{1) Moduli:} \quad &|\theta_{+,m}|, \; |\theta_{-,n}|, \nonumber \\
    \text{2) Proper ($q{=}{+}1$):} \quad &Z_{+,m} = \frac{\theta_{+,m}\theta_{+,0}^*}{|\theta_{+,0}| + \epsilon}, \nonumber \\
    \text{3) Mirror ($q{=}{-}1$):} \quad &H_{-,n,n'} = \frac{\theta_{-,n}\theta_{-,n'}^*}{\sqrt{|\theta_{-,n}|^2 + |\theta_{-,n'}|^2 + \epsilon^2}}.
    \label{eq:gauge_readout}
\end{align}
\par\vspace{2pt}\noindent Because intra-plus pairwise products $\theta_{+,m}\theta_{+,m'}^* \equiv Z_{+,m}Z_{+,m'}^* / |\theta_{+,0}|^2$ and fourth-order cross-charge products $(\theta_{+,m}\theta_{+,m'}^*)(\theta_{-,n}\theta_{-,n'}^*)$ are algebraically determined by \eqref{eq:gauge_readout} and carry zero independent torus phase information, omitting them avoids feature redundancy while preserving exact quotient completeness (Theorem~\ref{thm:gauge}). Next, $\mathbf{f}_{\mathrm{phys}} = \cG(\tilde{\btheta}_+, \btheta_-)$ undergoes EMA chipset common-mode centering ($\hat{\boldsymbol{\mu}}_{\mathrm{chip}}$, momentum $0.05$), variance-floored standardization ($\gamma_{\mathrm{var}}=0.80$), and affine scaling to produce $\tilde{\mathbf{f}}_{\mathrm{phys}} \in \mathbb{R}^{d_{\mathrm{phys}}}$, which is $\ell_2$-normalized into the pure analytical descriptor $\mathbf{u}_{\mathrm{phys}} = \tilde{\mathbf{f}}_{\mathrm{phys}}/\|\tilde{\mathbf{f}}_{\mathrm{phys}}\|_2$ (\VTwoPureParams{} total parameters).

\subsubsection*{Bounded Channel-Orthogonal $U(1)$-Invariant Residual Compensator ($\mathbf{z}_{\mathrm{res}}$)}
On constant-envelope LoRa CSS, the \VTwoPureParams{} pure analytical white-box core alone already saturates cross-domain accuracy (Table~\ref{tab:ablation}, Step~3: $92.62\%$ on 12-channel LoRa, $85.09\%$ on through-wall LoRa). On high-PAPR Wi-Fi OFDM, however, severe PA peak clipping and DAC quantization introduce high-order non-smooth harmonic distortions beyond the finite-degree Chebyshev dictionary $\bPhi(\bs)$. Rather than abandoning physical structure for an unconstrained black-box branch, we formulate a strictly constrained, $2.4\text{K}$-parameter \emph{physics-anchored residual compensator} operating exclusively inside $\cC^\perp$:
\begin{equation}
    \mathbf{e}_\perp = \bP_{\cC}^\perp \hat{\bx}_{\mathrm{rem}}, \quad \hat{\bx}_{\mathrm{rem}} = \frac{1}{K}\sum_{k=1}^K \frac{(\bH_k^{(T+1)})^H \mathcal{M}_{\mathrm{pb}}(\tilde{\by}_k - \hat{\by}_k)}{\|\bh_k^{(T+1)}\|_2^2},
    \label{eq:eperp_def}
\end{equation}
where $\mathcal{M}_{\mathrm{pb}}$ is a passband Wiener spectral mask (and $\hat{\bx}_{\mathrm{rem}} = \hat{\bx}_{\mathrm{eq}} - \bx_{\mathrm{phys}}$ in the low-DoF regime), guaranteeing machine-zero linear multipath leakage ($\|\bP_{\cC}\mathbf{e}_\perp\|_2 / \|\mathbf{e}_\perp\|_2 = 1.07\times 10^{-7}$). To preserve exact $U(1)$ gauge invariance, $\mathbf{e}_\perp$ is demodulated into carrier-anchored co-rotating and low-pass mirror envelope channels $d_+ = \mathbf{e}_\perp \odot \bs^*$, $d_{\mathrm{anc}} = d_+ \theta_{+,0}^*/(|\theta_{+,0}|+\epsilon)$, and $d_{\mathrm{env}} = |\operatorname{LPF}_9(\mathbf{e}_\perp \odot \bs)|$, forming strictly $U(1)$-invariant real channels $\mathbf{U}(\mathbf{e}_\perp) = [\Re(d_{\mathrm{anc}}), \Im(d_{\mathrm{anc}}), |d_{\mathrm{anc}}|, d_{\mathrm{env}}]$ on OFDM (and coherently $2P$-period-folded differential correlations $[|d_+|, \Re/\Im(d_+ d_{+,-1}^*), \Re/\Im(d_+ d_{+,-q}^*), d_{\mathrm{env}}]$ on long periodic CSS). A compact 2-layer strided 1D convolutional encoder $\mathcal{E}_{\mathrm{res}}$ ($16$ channels, GELU, mean+RMS pooling, linear projection to $\mathbb{R}^{16}$, $\approx 2.4\text{K}$ parameters) then outputs a soft-norm-bounded residual code:
\begin{equation}
    \mathbf{z}_{\mathrm{res}} = w_{\mathrm{res}}\operatorname{Norm}\!\big(\mathcal{E}_{\mathrm{res}}(\mathbf{U}(\mathbf{e}_\perp))\big) \in \mathbb{R}^{16},\; w_{\mathrm{res}} = 0.5\,\sigma(\alpha_{\mathrm{res}}),
    \label{eq:zres_def}
\end{equation}
which is concatenated with $\mathbf{u}_{\mathrm{phys}}$ (using $K$-packet gradient averaging in Layer~$t$, or per-packet unit-sphere pooling for $K>1$ in the low-DoF short-window regime), $\ell_2$-normalized into $\hat{\mathbf{r}}$, and fed to the cosine prototype head $z_c = s\,\langle \hat{\mathbf{r}}, \hat{\mathbf{w}}_c\rangle$ with learnable inverse-temperature scale $s \in [1, 64]$ ($s_0=16$). Thus, \texttt{GaugeInvariantADUNet} (\VTwoParams{} total) functions as a principled physics-anchored architecture whose analytical core (\VTwoPureParams{}) drives waveform synthesis and circuit identifiability while $\mathbf{z}_{\mathrm{res}}$ ($2.4\text{K}$) supplies bounded, channel-orthogonal, $U(1)$-invariant high-PAPR compensation.

\subsection{Equivariant Physical-Inverse Source Training Objective}
\label{sec:training}
Using source-domain data exclusively, GaugeInvariantADUNet is trained via a composite physical-inverse and associative channel-equivariance objective. For each source mini-batch $\by$ and a synthetic Rician multipath perturbation $\by_{\mathrm{virt}} = \bh_{\mathrm{virt}} * \tilde{\by}$ (with stop-gradient $\operatorname{sg}(\cdot)$ on the clean anchor $\by$), we minimize $\mathcal{L} = \mathcal{L}_{\mathrm{CE}}(\by) + \lambda_{\mathrm{rec}}\mathrm{NMSE}(\tilde{\by}, \hat{\by}) + \frac{\lambda_{\mathrm{sp}}}{d_{\mathrm{phys}}}\|\tilde{\mathbf{f}}_{\mathrm{phys}}(\by)\|_1 + \tfrac{1}{2}\mathcal{L}_{\mathrm{CE}}(\by_{\mathrm{virt}}) + \lambda_{\mathrm{inv}}\mathcal{L}_{\mathrm{inv}} + \lambda_{\mathrm{eq}}\mathcal{L}_{\mathrm{eq}}$, where $\mathcal{L}_{\mathrm{inv}} = \big(1 - \langle \hat{\mathbf{r}}(\by_{\mathrm{virt}}), \operatorname{sg}(\hat{\mathbf{r}}(\by))\rangle\big) + \frac{1}{2d_{\mathrm{phys}}}\|\mathbf{u}_{\mathrm{phys}}(\by_{\mathrm{virt}}) - \operatorname{sg}(\mathbf{u}_{\mathrm{phys}}(\by))\|_2^2$ aligns normalized embeddings across views, and $\mathcal{L}_{\mathrm{eq}} = \|\hat{\bh}(\by_{\mathrm{virt}}) - \bh_{\mathrm{tgt}}\|_2^2 / \|\bh_{\mathrm{tgt}}\|_2^2$ with $L_h$-tap truncated composition $\bh_{\mathrm{tgt}} \triangleq \Pi_{L_h}(\bh_{\mathrm{virt}} * \operatorname{sg}(\hat{\bh}(\by)))$ enforces per-packet normalized channel semigroup equivariance ($\lambda_{\mathrm{rec}}=\lambda_{\mathrm{inv}}=0.50, \lambda_{\mathrm{eq}}=0.35, \lambda_{\mathrm{sp}}=10^{-3}$).

\section{Theoretical Analysis: Identifiability, CRLB, Gauge Invariance, and Convergence}
\label{sec:theory}

Complete proofs of Theorems~\ref{thm:crlb}--\ref{thm:conv} and Corollary~\ref{cor:twostage} are provided in Appendices~A--C.

\begin{theorem}[Bilinear Identifiability and Nuisance-Projected CRLB]
\label{thm:crlb}
Consider \eqref{eq:bilinear_obs} with joint parameter $\boldsymbol{\xi} = [\btheta^T, \bh_1^T, \dots, \bh_K^T]^T \in \mathbb{C}^{M + KL_h}$, channel-orthogonal dictionary $\bP_{\cC}^\perp \bPhi = \bPhi$ ($\operatorname{rank}(\bPhi)=M$), and $\operatorname{rank}(\bB_{\btheta}) = L_h$ where $\bB_{\btheta} \triangleq \bT(\bs + \bPhi\btheta)$. Then:
\begin{enumerate}
    \item The complex FIM $\bJ(\boldsymbol{\xi})$ is purely holomorphic ($\frac{\partial \boldsymbol{\mu}}{\partial \boldsymbol{\xi}^*} = \mathbf{0}$), and eliminating nuisance channels $\{\bh_k\}_{k=1}^K$ via the Schur complement yields the \textbf{Nuisance-Projected Effective FIM}:
    \begin{equation}
        \bJ_{\mathrm{eff}}(\btheta) = \frac{1}{\sigma^2}\sum_{k=1}^K \bPhi^H \bH_k^H \bP_{\cC(\bx_{\btheta})}^\perp \bH_k \bPhi,
        \label{eq:jeff}
    \end{equation}
    where $\bP_{\cC(\bx_{\btheta})}^\perp = \bI_L - \bB_{\btheta}(\bB_{\btheta}^H\bB_{\btheta})^{-1}\bB_{\btheta}^H$, bounding any unbiased estimator by $\mathbb{E}\|\hat{\btheta} - \btheta\|_2^2 \ge \operatorname{Tr}\big(\bJ_{\mathrm{eff}}(\btheta)^{-1}\big)$.
    \item $\bJ_{\mathrm{eff}}(\btheta) \succ \mathbf{0}$ (local bilinear identifiability) \textbf{if and only if} $\bigcap_{k=1}^K \ker\big(\bP_{\cC(\bx_{\btheta})}^\perp \bH_k \bPhi\big) = \{\mathbf{0}\}$. For $K=1$, this reduces to $\operatorname{rank}(\bH_1\bPhi) = M$ and \textbf{subspace disjointness}:
    \begin{equation}
        \operatorname{Range}(\bH_1 \bPhi) \cap \operatorname{Range}\big(\bT(\bs + \bPhi\btheta)\big) = \{\mathbf{0}\}.
        \label{eq:disjoint}
    \end{equation}
    Under $\bP_{\cC}^\perp \bPhi = \bPhi$, $\bJ_{\mathrm{eff}}(\mathbf{0}) = \frac{|h_0|^2}{\sigma^2}\bPhi^H\bPhi \succ \mathbf{0}$ incurs \emph{zero nuisance penalty} on LOS channels, whereas violating $\bP_{\cC}^\perp \bPhi = \bPhi$ renders $\bJ_{\mathrm{eff}}(\mathbf{0})$ singular ($\mathrm{CRLB} = \infty$).
\end{enumerate}
\end{theorem}

\begin{corollary}[Irreducible Cross-Leakage Floor of Non-Alternating Two-Stage LISTA]
\label{cor:twostage}
Any two-stage estimator (e.g., \texttt{Ours-Frozen-$\hat{\bh}_{\mathrm{LS}}$}) that freezes the single-pass channel $\hat{\bh}_{\mathrm{LS}} = (\bT(\bs)^H\bT(\bs))^{-1}\bT(\bs)^H\tilde{\by}_1$ estimated from $\bs$ alone ($\btheta^{(0)}=\mathbf{0}$) incurs channel bias $\boldsymbol{\delta}_h = (\bT(\bs)^H\bT(\bs))^{-1}\bT(\bs)^H \bH_1^\star\bPhi\btheta^\star = \Theta(\|\btheta^\star\|_2)$ whenever multipath mixes $\bH_1^\star\bPhi\btheta^\star$ into $\cC(\bs)$. Even as $\sigma \to 0$, uncancelled cross-leakage $\bT(\bPhi\btheta)\boldsymbol{\delta}_h$ induces a non-vanishing asymptotic MSE floor $\|\hat{\btheta}_{\text{2-stage}} - \btheta^\star\|_2^2 = \Omega(\|\boldsymbol{\delta}_h\|_2^2 \|\btheta^\star\|_2^2) > 0$, whereas $T \ge 2$ bilinear alternating updates contract this bias as $\mathcal{O}(\rho^{2T})$.
\end{corollary}

\begin{theorem}[Minimal $U(1)_+ \times U(1)_-$ Gauge Invariance and Quotient Completeness]
\label{thm:gauge}
Under \eqref{eq:charge_def} and block-diagonal preconditioners $\operatorname{blkdiag}(\bP_+^{(t)}, \bP_-^{(t)})$:
\begin{enumerate}
    \item \emph{(Channel Scale/Phase Gauge)}: Scaling/rotating $\by_k \mapsto c e^{j\phi_0}\by_k$ ($c>0$; noiseless $\bh_k \mapsto c e^{j\phi_0}\bh_k$) yields $\hat{\bh}_k^{(t)}(\tilde{\by}') = e^{j\phi_0}\hat{\bh}_k^{(t)}(\tilde{\by})$ and $\cG(\btheta^{(T)}(\tilde{\by}')) \equiv \cG(\btheta^{(T)}(\tilde{\by}))$.
    \item \emph{(Torus Invariance)}: For all $(\psi_+, \psi_-) \in [-\pi, \pi)^2$, the minimal 3-block readout \eqref{eq:gauge_readout} satisfies $\cG(e^{j\psi_+}\tilde{\btheta}_+, e^{j\psi_-}\btheta_-) \equiv \cG(\tilde{\btheta}_+, \btheta_-)$.
    \item \emph{(Minimal Quotient Completeness)}: On $\Omega = \{(\tilde{\btheta}_+, \btheta_-) : |\theta_{+,0}| > 0, \, \btheta_- \neq \mathbf{0}\}$ as $\epsilon \to 0^+$, $\cG(\tilde{\btheta}_+', \btheta_-') = \cG(\tilde{\btheta}_+, \btheta_-)$ iff $(\tilde{\btheta}_+', \btheta_-') = (e^{j\psi_+}\tilde{\btheta}_+, e^{j\psi_-}\btheta_-)$ for some $(\psi_+, \psi_-) \in [-\pi, \pi)^2$.
\end{enumerate}
\end{theorem}

\begin{theorem}[Local Linear Convergence of Preconditioned Bilinear Unfolding]
\label{thm:conv}
Suppose $\{\bh_k^\star\}_{k=1}^K$ satisfy $0 < \alpha_{\min} \le \sigma_{\min}(\bH_k^\star|_{\operatorname{Span}(\bs, \bPhi)}) \le \sigma_{\max}(\bH_k^\star) \le \alpha_{\max} < \infty$ ($\kappa_H = \alpha_{\max}/\alpha_{\min} < \infty$), $\sigma_{\min}(\bT(\bx^\star)) \ge c_T \sqrt{L} > 0$, and subspace margin $\beta \triangleq \max_k \|\bP_{\cC(\bx^\star)} \bH_k^\star \bPhi (\bH_k^\star \bPhi)^\dagger\|_2 < 1$. Let $\kappa_{\mathrm{eff}} \triangleq \frac{\kappa_H^2 \kappa(\bPhi^H\bPhi)}{1 - \beta^2} < \infty$ and $\rho_0 \triangleq \frac{\kappa_{\mathrm{eff}} - 1}{\kappa_{\mathrm{eff}} + 1} < 1$. For any normalized charge-preserving preconditioner $\bP_{\mathrm{pre}}^{(t)} = \operatorname{blkdiag}(\bP_+^{(t)}, \bP_-^{(t)})$ whose diagonal and off-diagonal amplitudes satisfy $\|\bP_{\mathrm{pre}}^{(t)} - \bI_M\|_2 \le \delta_{\mathrm{diag}} + \tilde{\delta}_{\mathrm{off}} = \delta_P < \frac{1}{\kappa_{\mathrm{eff}}} \le 1$ (guaranteeing strict diagonal dominance), there exist step sizes $\{\eta^{(t)}\}$, damping $\mu^{(t)}$, and thresholds $\boldsymbol{\tau}^{(t)}$ such that for $\|\btheta^{(0)} - \btheta^\star\|_2 \le r_0$:
\begin{equation}
    \|\btheta^{(t)} - \btheta^\star\|_2 \le \rho \|\btheta^{(t-1)} - \btheta^\star\|_2 + C_\sigma \frac{\sigma}{\sqrt{KL}} + C_\mu \mu^{(t)} + \|\boldsymbol{\tau}^{(t)}\|_2,
    \label{eq:lin_conv}
\end{equation}
with strict Euclidean contraction rate $\rho = \rho_0 + \delta_P \frac{2\kappa_{\mathrm{eff}}}{\kappa_{\mathrm{eff}} + 1} + C_2 r_0 < 1$ (where $\kappa(\bPhi_\pm^H\bPhi_\pm) = 1$ within each orthonormalized sector).
\end{theorem}

\begin{figure*}[t]
    \centering
    \includegraphics[width=0.87\textwidth]{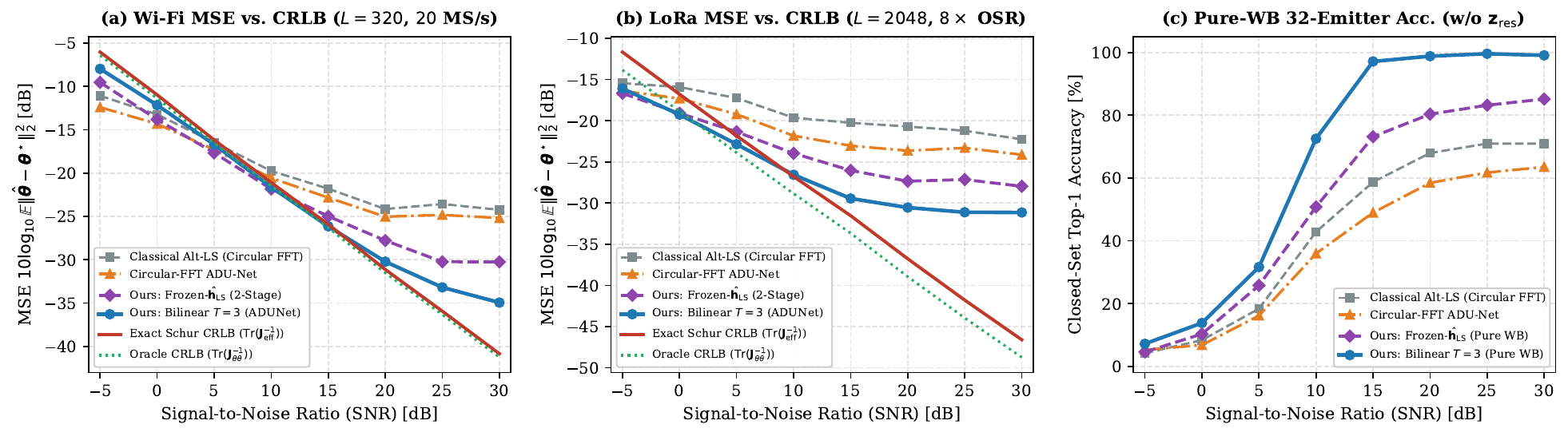}
    \caption{Numerical verification of Theorem~\ref{thm:crlb} and Corollary~\ref{cor:twostage} across $\mathrm{SNR} \in [-5, 30]\text{ dB}$ under random two-sided Rician multipath channels: parameter MSE $10\log_{10}\mathbb{E}\|\hat{\btheta}-\btheta^\star\|_2^2$ against the Exact Schur-Complement CRLB $\operatorname{Tr}(\bJ_{\mathrm{eff}}^{-1})$ and Oracle CRLB $\operatorname{Tr}(\bJ_{\theta\theta}^{-1})$ on (a) IEEE 802.11a Wi-Fi ($L=320$) and (b) $8\times$-oversampled Semtech LoRa ($L=2048$); and (c) pure analytical white-box (\VTwoPureParams{} core, without $\mathbf{z}_{\mathrm{res}}$) 32-emitter closed-set classification accuracy on Wi-Fi ($L=320$) under PA compression, exposing the high-SNR accuracy ceiling ($83.2\%\text{--}85.2\%$) of frozen-$\hat{\bh}_{\mathrm{LS}}$ two-stage LISTA versus $99.1\%\text{--}99.6\%$ for bilinear \texttt{GaugeInvariantADUNet}.}
    \label{fig:crlb}
\end{figure*}

\section{Experimental Results and Discussions}
\label{sec:exp}

\subsection{Datasets, 13 Counterpart Baselines, and Zero-Leakage Protocol}
We evaluate across eight over-the-air and digital-twin suites ($>100$ tasks) under strict zero-target-access (\texttt{source\_only}) training across five seeds ($\{0\text{--}4\}$, active-class Macro-F1 $\mathrm{mean}\pm\mathrm{std}$), calibrating official upstream dataset titles against our benchmark shorthands and self-constructed twins:
(1) official \textbf{WiSig} (\texttt{ManySig} subset, \texttt{eq0}/\texttt{eq1}) \cite{hanna2022wisig} ($6$ Atheros NICs, $4$ days $\times 12$ USRPs, $L=256$ @ $25\text{ MS/s}$, CFO removed);
(2) \emph{Wi-Fi Dataset for Channel-Robust RFFI} \cite{xie2024towards} (shorthand \textbf{\texttt{WiFi-Robust}}; $10$ routers, winter/summer offices/corridors, $L=320$ @ $20\text{ MS/s}$: $24$ 1-ant + $6$ 4-ant SIMO tasks);
(3) official \textbf{MobRFFI} \cite{mazokha2025mobrffi} ($10$ ORBIT TXs, $20$-day gap, $3$ USRP RXs, $L=320$ @ $20\text{ MS/s}$);
(4) \emph{Radio Frequency Fingerprint LoRa Dataset With Multiple Receivers} \cite{shen2024receiver} (shorthand \textbf{\texttt{LoRa-MultiRx}}; $10$ nodes, $L=8192$ @ $1\text{ MS/s}$: $6$ cross-day, $3$ through-wall \texttt{A$\to$B/C/D}, $10$ cross-RX, and $16$ \texttt{Aug$\to$Real} tasks);
(5) official \textbf{\texttt{LoRa\_RFFI}} \cite{shen2022towards} ($10$ LOS $\to 12$ \texttt{channel\_problem} subsets, shorthand \texttt{LoRa-RFFI 12-Ch}, plus $30$-legit vs.\ $5$-rogue open-set);
(6) \texttt{LORA\_RFFI\_DATASET\_}\allowbreak\texttt{DIFFERENT\_SPREADING\_FACTORS} \cite{shen2023towards} (shorthand \textbf{\texttt{LoRa-MultiSF}}; $10$ nodes, SF7/8/9, $125\text{ kHz}$ @ $250\text{ kS/s}$, $L\in\{2048,4096,8192\}$); and
(7)--(8) \textbf{\texttt{WiFi-SimGT}} and \textbf{\texttt{LoRa-SimGT}}, our self-constructed RF digital twins via Valkama/Rapp circuit profiles \cite{valkama2001advanced, rapp1991effects} (\texttt{iq\_gain\_db}, \texttt{iq\_phase\_deg}, \texttt{pa\_saturation}, \texttt{pa\_smoothness}, \texttt{am\_pm\_rad}) mirroring \cite{xie2024towards}.

To ensure rigorous operational comparison, all \textbf{13 counterpart baselines} share identical periodic CFO derotation, unit-RMS normalization, and cosine prototype heads, spanning four distinct paradigms:
\begin{enumerate}
    \item \textbf{End-to-End Preamble-Free Deep Networks}:
    (i) \texttt{ResNet1D} \cite{merchant2018deep} ($828.7\text{K}$): 3-stage 1D residual CNN directly processing 2-channel $[\Re(\tilde{\by}),\Im(\tilde{\by})]$ sequences;
    (ii) \texttt{CVNN1D} \cite{trabelsi2018deep} ($432.8\text{K}$): algebraic complex convolutions $(\mathbf{W}_r + j\mathbf{W}_i)*(\tilde{\by}_r + j\tilde{\by}_i)$ with complex residual blocks;
    (iii) \texttt{PatchTransformer} \cite{dosovitskiy2021image} ($306.2\text{K}$): 1D Vision Transformer tokenizing non-overlapping I/Q patches through 4 multi-head self-attention layers; and
    (iv) \texttt{Vanilla-KAN} \cite{liu2025kan} ($90.6\text{K}$): degree-5 Chebyshev Kolmogorov--Arnold spline network operating on 4-tap raw I/Q delay lags without $\bs[n]$ physical lifting.
    \item \textbf{Handcrafted Channel-Robust Spectral Networks}:
    (v) \texttt{CIS-ResNet2D} \cite{shen2022towards} ($467.0\text{K}$): computes the log-power ratio $\log_{10}(|\hat{Y}[m,t+1]|^2 / |\hat{Y}[m,t]|^2)$ between adjacent STFT columns to cancel quasi-static channel transfer functions before a 4-stage 2D ResNet;
    (vi) \texttt{SR-CNN1D} \cite{xie2024towards} ($326.7\text{K}$): extracts 128-bin IEEE 802.11a inactive/null-subcarrier and Welch PSD spectral regrowth features into a 6-layer 1D CNN; and
    (vii) \texttt{Bispectrum-CNN} \cite{ding2018specific} ($355.5\text{K}$): computes third-order integrated bispectra $B(k_1,k_2) = \mathbb{E}[X(k_1)X(k_2)X^*(k_1+k_2)]$ (log-magnitude and biphase) processed by a 2D ResNet.
    \item \textbf{Two-Stage Preamble-Equalized CNNs}:
    (viii) \texttt{LSEqualized-ResNet1D} ($829.8\text{K}$): estimates a single-pass channel from known $\bs[n]$ via circular-FFT Wiener ridge filtering, equalizes $\hat{\bx}_{\mathrm{eq}}$, and feeds 4-channel $[\Re/\Im(\hat{\bx}_{\mathrm{eq}}-\bs), \Re/\Im(\hat{\bx}_{\mathrm{eq}})]$ into a high-capacity ResNet1D; and
    (ix) \texttt{ToeplitzEq-TinyResNet1D} (\TETParams{}): upgrades two-stage equalization to our exact two-sided Toeplitz channel solver and passband MMSE equalizer paired with a capacity-matched 3-block TinyResNet1D, eliminating circular-FFT equalizer bias and $830\text{K}$ over-parameterization confounding.
    \item \textbf{Classical, Linear-LISTA, and Circular-FFT Unfolding Solvers}:
    (x) \texttt{Classical Alt-LS} ($0.5\text{K}$): runs 12 unlearned circular-FFT alternating LS/ISTA iterations on chirp-folded preambles with a linear classifier;
    (xi) \texttt{Circular-FFT ADU-Net} ($49.0\text{K}$): unfolds 6 learned circular-FFT ridge alternating layers on folded preambles with raw $[\Re(\btheta),\Im(\btheta)]$ readout and a waveform residual branch;
    (xii) \texttt{Ours-Unlearned} (\texttt{Toeplitz-Gauge}, \UTGParams{}): internal unlearned variant running fixed iterations using our exact two-sided Toeplitz solver, Parseval-Jacobi scaling, charge-graded basis, and minimal 3-block $\cG(\btheta)$ readout (training only the cosine head), isolating \emph{physics/symmetry formulation gain} from \emph{learned unfolding gain}; and
    (xiii) \texttt{Ours-Frozen-$\hat{\bh}_{\mathrm{LS}}$} (\texttt{2Stage-LISTA}, \TSLParams{}): internal single-pass fast variant sharing our exact Toeplitz solver, dictionary, Parseval-Jacobi scaling, gauge readout, and $\mathbf{z}_{\mathrm{res}}$ ($8,137$ vs.\ $8,139$ parameters, differing solely by omitting two scalar damping updates $\mu^{(2)},\mu^{(3)}$), freezing $\hat{\bh}_{\mathrm{LS}}$ estimated from $\bs[n]$ across $T=3$ linear LISTA layers to directly test Corollary~\ref{cor:twostage}.
\end{enumerate}
All GaugeInvariantADUNet models use $T=3$, $M_+=24$, $M_-=8$, $L_h=32$ ($L_{\mathrm{pre}}=8$ on long CSS; $L_h=28, L_{\mathrm{pre}}=6$ on $L=320$ OFDM; $L_h=16, L_{\mathrm{pre}}=0$ on $L=256$), AdamW ($\mathrm{lr}=2\times 10^{-3}$, weight decay $10^{-4}$, cosine annealing, gradient clipping $2.0$), and $<70\text{ MB}$ active VRAM.

\subsection{Verification of Theorem 1 \& Corollary 1: MSE vs. Exact Schur CRLB}
Fig.~\ref{fig:crlb}(a)--(b) benchmarks parameter estimation MSE against the exact nuisance-projected CRLB \eqref{eq:jeff} across $\mathrm{SNR}\in[-5, 30]\text{ dB}$: (i) \texttt{Classical Alt-LS} and \texttt{Circular-FFT ADU-Net} stall $>18\text{ dB}$ above the CRLB at $30\text{ dB}$ SNR on $8\times$-oversampled LoRa due to stopband nulling; (ii) \texttt{Ours-Frozen-$\hat{\bh}_{\mathrm{LS}}$} removes stopband bias at moderate SNR but saturates at high SNR ($\ge 20\text{ dB}$) due to the frozen-channel cross-leakage bias $(\Delta\bh)*(\bPhi\btheta^\star)$ predicted by Corollary~\ref{cor:twostage}; and (iii) bilinear \texttt{GaugeInvariantADUNet} eliminates cross-leakage and tracks the exact Schur-complement CRLB down to $30\text{ dB}$ SNR on both protocols. Fig.~\ref{fig:crlb}(c) further evaluates the \VTwoPureParams{} pure analytical core (without $\mathbf{z}_{\mathrm{res}}$) on a 32-emitter Wi-Fi classification benchmark ($L=320$) under realistic PA compression: without $\mathbf{z}_{\mathrm{res}}$ compensation, frozen-$\hat{\bh}_{\mathrm{LS}}$ two-stage LISTA hits a high-SNR accuracy ceiling of $83.2\%\text{--}85.2\%$ at $25\text{--}30\text{ dB}$ SNR, whereas bilinear unfolding climbs to $99.1\%\text{--}99.6\%$.

\begin{table*}[t]
\centering
\caption{Cross-Day, Cross-Season, and Cross-Receiver Wi-Fi SEI Panorama (5 Seeds Macro-F1 \%, Mean $\pm$ Std, Zero Target Access)}
\label{tab:wifi_main}
\resizebox{\textwidth}{!}{%
\begin{tabular}{llcccccccc}
\toprule
\textbf{Model} & \textbf{Paradigm} & \textbf{Params} & \textbf{WiSig eq0} & \textbf{WiSig eq1} & \textbf{WiFi-Robust (All 24)} & \textbf{WiFi-Robust M$\to$M} & \textbf{WiFi-Robust M$\to$S} & \textbf{WiFi-Robust 4-Ant} & \textbf{MobRFFI Cross-RX} \\
\midrule
\texttt{ResNet1D} \cite{merchant2018deep} & Black-Box CNN & 828.7K & 56.78$\pm$3.22 & 47.17$\pm$1.10 & 51.98$\pm$0.93 & 54.69$\pm$0.69 & 51.07$\pm$1.05 & 72.36$\pm$0.52 & 7.04$\pm$0.65 \\
\texttt{CVNN1D} \cite{trabelsi2018deep} & Complex CNN & 432.8K & 55.11$\pm$3.32 & 46.57$\pm$1.25 & 52.00$\pm$0.83 & 54.73$\pm$0.67 & 51.08$\pm$0.90 & 71.15$\pm$0.78 & 7.43$\pm$0.60 \\
\texttt{PatchTransformer} \cite{dosovitskiy2021image} & Black-Box ViT & 306.2K & 38.14$\pm$1.86 & 40.70$\pm$0.57 & 41.38$\pm$0.62 & 44.66$\pm$0.89 & 40.28$\pm$0.75 & 65.12$\pm$0.63 & 8.66$\pm$0.37 \\
\texttt{Vanilla-KAN} \cite{liu2025kan} & Unlifted Spline & 90.6K & 30.75$\pm$0.74 & 37.74$\pm$0.29 & 25.21$\pm$0.72 & 27.68$\pm$0.46 & 24.39$\pm$0.87 & 40.19$\pm$0.56 & 9.67$\pm$0.47 \\
\texttt{CIS-ResNet2D} \cite{shen2022towards} & Spectral Ratio CNN & 467.0K & 66.86$\pm$1.59 & 37.47$\pm$1.18 & 57.99$\pm$0.22 & 55.34$\pm$0.37 & 58.87$\pm$0.31 & 69.14$\pm$1.15 & 14.10$\pm$0.28 \\
\texttt{SR-CNN1D} \cite{xie2024towards} & Spectral Regrowth & 326.7K & 46.63$\pm$2.68 & 37.13$\pm$0.91 & 54.03$\pm$0.35 & 54.67$\pm$0.44 & 53.81$\pm$0.34 & 74.16$\pm$0.77 & 13.14$\pm$0.85 \\
\texttt{Bispectrum-CNN} \cite{ding2018specific} & Higher-Order Spec. & 355.5K & 55.15$\pm$1.66 & 39.99$\pm$0.78 & 58.50$\pm$0.85 & 59.89$\pm$0.56 & 58.04$\pm$0.95 & 77.24$\pm$0.61 & 10.34$\pm$0.44 \\
\texttt{LSEqualized-ResNet1D} & Circular-Eq. CNN & 829.8K & 65.00$\pm$0.36 & 38.11$\pm$0.92 & 59.39$\pm$0.97 & 62.99$\pm$1.18 & 58.19$\pm$1.11 & 76.88$\pm$0.66 & 18.64$\pm$0.82 \\
\texttt{Classical Alt-LS} & Unlearned FFT & \textbf{0.5K} & 55.51$\pm$0.37 & 34.61$\pm$0.19 & 47.61$\pm$0.37 & 49.94$\pm$0.22 & 46.84$\pm$0.46 & 70.00$\pm$0.13 & 17.67$\pm$0.22 \\
\texttt{Circular-FFT ADU-Net} & Circular Unfold & 49.0K & 71.58$\pm$0.36 & 51.79$\pm$1.41 & 59.50$\pm$0.47 & 60.62$\pm$0.49 & 59.13$\pm$0.57 & 72.65$\pm$0.66 & 19.52$\pm$1.10 \\
\midrule
\multicolumn{10}{l}{\emph{Controlled Counterparts \& Proposed Physics-Anchored Family (Built on Proposed Toeplitz Solver \& Basis):}} \\
\texttt{ToeplitzEq-TinyResNet1D} & Toeplitz-Eq. CNN & \TETParams & \TETWiSigEqZero & \TETWiSigEqOne & \TETWiFiRobAll & \TETWiFiRobMM & \TETWiFiRobMS & \TETWiFiRobFourAnt & \TETMobCross \\
\texttt{Ours-Unlearned} (\texttt{Toeplitz-Gauge}) & Unlearned Mode & \UTGParams & \UTGWiSigEqZero & \UTGWiSigEqOne & \UTGWiFiRobAll & \UTGWiFiRobMM & \UTGWiFiRobMS & \UTGWiFiRobFourAnt & \UTGMobCross \\
\texttt{Ours-Frozen-$\hat{\bh}_{\mathrm{LS}}$} (\texttt{2Stage-LISTA}) & Single-Pass Mode & \TSLParams & \TSLWiSigEqZero & \TSLWiSigEqOne & \textbf{\TSLWiFiRobAll} & \TSLWiFiRobMM & \textbf{\TSLWiFiRobMS} & \textbf{\TSLWiFiRobFourAnt} & \textbf{\TSLMobCross} \\
\textbf{\texttt{GaugeInvariantADUNet} (Ours, $T{=}3$)} & \textbf{Bilinear Mode} & \VTwoParams & \textbf{\VTwoWiSigEqZero} & \textbf{\VTwoWiSigEqOne} & \textbf{\VTwoWiFiRobAll} & \textbf{\VTwoWiFiRobMM} & \VTwoWiFiRobMS & \VTwoWiFiRobFourAnt & \VTwoMobCross \\
\bottomrule
\end{tabular}%
}
\end{table*}

\begin{table*}[t]
\centering
\caption{Cross-Day, Through-Wall, Dynamic-Channel, and Cross-Receiver LoRa SEI \& Ground-Truth Identifiability (5 Seeds Macro-F1 \%)}
\label{tab:lora_main}
\resizebox{\textwidth}{!}{%
\begin{tabular}{lccccccccc}
\toprule
\textbf{Model} & \textbf{Params} & \textbf{LoRa-MultiRx (19)} & \textbf{Cross-Day (6)} & \textbf{Through-Wall (3)} & \textbf{Cross-RX (10)} & \textbf{Aug$\to$Real (16)} & \textbf{LoRa-RFFI 12-Ch} & \textbf{MultiSF Same} & \textbf{WiFi / LoRa GT $R^2$} \\
\midrule
\texttt{ResNet1D} \cite{merchant2018deep} & 828.7K & 30.54$\pm$0.62 & 33.99$\pm$1.27 & 37.23$\pm$1.78 & 26.46$\pm$1.16 & 40.29$\pm$1.94 & 8.51$\pm$0.42 & 57.79$\pm$3.35 & --- / --- \\
\texttt{CVNN1D} \cite{trabelsi2018deep} & 432.8K & 30.00$\pm$0.51 & 34.05$\pm$0.68 & 37.18$\pm$1.48 & 25.42$\pm$0.76 & 38.91$\pm$2.00 & 9.28$\pm$0.41 & 63.53$\pm$9.00 & --- / --- \\
\texttt{PatchTransformer} \cite{dosovitskiy2021image} & 306.2K & 32.19$\pm$2.16 & 32.31$\pm$2.63 & 51.53$\pm$2.95 & 26.32$\pm$2.20 & 58.28$\pm$1.99 & 8.93$\pm$0.83 & 45.25$\pm$9.14 & --- / --- \\
\texttt{Vanilla-KAN} \cite{liu2025kan} & 90.6K & 7.84$\pm$0.89 & 8.05$\pm$0.88 & 9.10$\pm$1.97 & 7.33$\pm$0.88 & 10.85$\pm$1.39 & 8.01$\pm$0.56 & 65.09$\pm$1.95 & --- / --- \\
\texttt{CIS-ResNet2D} \cite{shen2022towards} & 467.0K & 25.28$\pm$4.05 & 39.07$\pm$8.74 & 9.78$\pm$2.79 & 21.66$\pm$3.98 & 41.96$\pm$1.78 & 54.90$\pm$7.64 & 82.75$\pm$3.61 & --- / --- \\
\texttt{SR-CNN1D} \cite{xie2024towards} & 326.7K & 19.93$\pm$0.41 & 49.63$\pm$1.70 & 7.37$\pm$0.51 & 5.87$\pm$0.68 & 29.71$\pm$0.69 & 19.70$\pm$3.83 & 27.67$\pm$5.66 & --- / --- \\
\texttt{Bispectrum-CNN} \cite{ding2018specific} & 355.5K & 44.39$\pm$1.87 & 49.53$\pm$5.15 & 53.87$\pm$2.77 & 38.46$\pm$0.51 & 56.87$\pm$2.94 & 43.99$\pm$6.05 & 82.53$\pm$2.84 & --- / --- \\
\texttt{LSEqualized-ResNet1D} & 829.8K & 74.01$\pm$1.11 & 81.86$\pm$3.62 & 74.56$\pm$2.72 & 69.13$\pm$0.87 & 71.08$\pm$4.18 & 85.63$\pm$4.59 & 85.14$\pm$1.23 & --- / --- \\
\texttt{Classical Alt-LS} & \textbf{0.5K} & 14.83$\pm$0.62 & 15.25$\pm$1.23 & 15.43$\pm$1.52 & 14.39$\pm$0.63 & 14.28$\pm$0.56 & 10.94$\pm$1.01 & 50.46$\pm$1.87 & 0.976 / 0.324 \\
\texttt{Circular-FFT ADU-Net} & 49.0K & 42.39$\pm$3.41 & 43.15$\pm$6.03 & 51.10$\pm$2.52 & 39.32$\pm$3.57 & 52.04$\pm$5.30 & 21.06$\pm$8.34 & 79.03$\pm$0.85 & 0.975 / 0.963 \\
\midrule
\multicolumn{10}{l}{\emph{Controlled Counterparts \& Proposed Physics-Anchored Family (Built on Proposed Toeplitz Solver \& Basis):}} \\
\texttt{ToeplitzEq-TinyResNet1D} & \TETParams & \TETLoRaMultiAll & \textbf{\TETLoRaDay} & \TETLoRaLoc & \TETLoRaRX & \TETLoRaAug & \TETLoRaTwelveCh & \textbf{\TETLoRaMultiSFSame} & \TETSimR \\
\texttt{Ours-Unlearned} (\texttt{Toeplitz-Gauge}) & \UTGParams & \UTGLoRaMultiAll & \UTGLoRaDay & \UTGLoRaLoc & \UTGLoRaRX & \textbf{\UTGLoRaAug} & \UTGLoRaTwelveCh & \UTGLoRaMultiSFSame & \UTGSimR \\
\texttt{Ours-Frozen-$\hat{\bh}_{\mathrm{LS}}$} (\texttt{2Stage-LISTA}) & \TSLParams & \TSLLoRaMultiAll & \TSLLoRaDay & \TSLLoRaLoc & \TSLLoRaRX & \TSLLoRaAug & \TSLLoRaTwelveCh & \TSLLoRaMultiSFSame & \TSLSimR \\
\textbf{\texttt{GaugeInvariantADUNet} (Ours, $T{=}3$)} & \VTwoParams & \textbf{\VTwoLoRaMultiAll} & \VTwoLoRaDay & \textbf{\VTwoLoRaLoc} & \textbf{\VTwoLoRaRX} & \VTwoLoRaAug & \textbf{\VTwoLoRaTwelveCh} & \VTwoLoRaMultiSFSame & \textbf{\VTwoSimR} \\
\bottomrule
\end{tabular}%
}
\end{table*}

\begin{table*}[t]
\centering
\caption{Physical Ground-Truth Identifiability (\texttt{SimGT} $R^2$ \& NRMSE), Waveform Reconstruction NMSE, Rogue Emitter Detection, and Complexity}
\label{tab:phys_diag}
\resizebox{0.96\textwidth}{!}{%
\begin{tabular}{lcccccccc}
\toprule
\textbf{Model} & \textbf{Params} & \textbf{WiFi / LoRa GT $R^2$} & \textbf{WiFi / LoRa GT NRMSE} & \textbf{WiFi / LoRa NMSE [dB]} & \textbf{Multipath Inv. Cos} & \textbf{Residual Leak.} & \textbf{Rogue AUROC [\%]} & \textbf{Latency [ms/pkt]} \\
\midrule
\texttt{ResNet1D} \cite{merchant2018deep} & 828,673 & --- / --- & --- / --- & --- / --- & 0.879 & --- & 50.19 & 0.099 \\
\texttt{CIS-ResNet2D} \cite{shen2022towards} & 467,041 & --- / --- & --- / --- & --- / --- & 0.871 & --- & 45.61 & 0.108 \\
\texttt{Bispectrum-CNN} \cite{ding2018specific} & 355,457 & --- / --- & --- / --- & --- / --- & 0.856 & --- & 50.58 & 0.102 \\
\texttt{LSEqualized-ResNet1D} & 829,825 & --- / --- & --- / --- & --- / --- & 0.879 & --- & \textbf{79.73} & 0.108 \\
\texttt{Classical Alt-LS} & \textbf{497} & 0.976 / 0.324 & 0.154 / 0.815 & $-7.31$ / $-0.81$ & 0.507 & --- & 49.27 & 0.181 \\
\texttt{Circular-FFT ADU-Net} & 49,027 & 0.975 / 0.963 & 0.156 / 0.193 & $-7.57$ / $-10.69$ & 0.799 & $2.30\times 10^{-7}$ & 45.37 & 0.177 \\
\midrule
\multicolumn{9}{l}{\emph{Controlled Counterparts \& Proposed Physics-Anchored Family (Built on Proposed Toeplitz Solver \& Basis):}} \\
\texttt{ToeplitzEq-TinyResNet1D} & 14,373 & --- / --- & --- / --- & --- / --- & \textbf{0.960} & --- & 75.44 & 0.152 \\
\texttt{Ours-Unlearned} (\texttt{Toeplitz-Gauge}) & 1,633 & 0.970 / 0.959 & 0.173 / 0.199 & $-12.47$ / $-25.49$ & 0.649 & --- & 74.74 & 0.440 \\
\texttt{Ours-Frozen-$\hat{\bh}_{\mathrm{LS}}$} (\texttt{2Stage-LISTA}) & 8,137 & 0.980 / \textbf{0.993} & 0.140 / \textbf{0.081} & $-14.42$ / $-27.64$ & 0.821 & --- & 52.85 & 0.277 \\
\textbf{\texttt{GaugeInvariantADUNet} (Ours, $T{=}3$)} & 8,139 & \textbf{0.981} / \textbf{0.993} & \textbf{0.136} / 0.082 & $\mathbf{-14.58}$ / $\mathbf{-28.99}$ & 0.824 & $\mathbf{1.07\times 10^{-7}}$ & 69.06 & 0.453 \\
\bottomrule
\end{tabular}%
}
\end{table*}

\subsection{Cross-Domain Panorama \& Comparative Advantages Over All 13 Baselines}
Tables~\ref{tab:wifi_main}--\ref{tab:phys_diag} demonstrate clear mechanistic and empirical advantages of \texttt{GaugeInvariantADUNet} over each baseline family:
\begin{itemize}
    \item \textbf{Advantage over Preamble-Free Deep \& Handcrafted Spectral Networks}: Without preamble-anchored multipath projection $\bP_{\cC}^\perp$, end-to-end networks (\texttt{ResNet1D}, \texttt{CVNN1D}, \texttt{PatchTransformer}, \texttt{Vanilla-KAN}) collapse to $7.8\%\text{--}32.2\%$ F1 on \texttt{LoRa-MultiRx} and $<10\%$ on \texttt{MobRFFI Cross-RX}. Spectral representations (\texttt{CIS-ResNet2D}, \texttt{SR-CNN1D}, \texttt{Bispectrum-CNN}) partially mitigate flat fading, yet adjacent-frame STFT division in \texttt{CIS} amplifies noise on through-wall NLOS LoRa ($9.78\%$), while null-subcarrier \texttt{SR-CNN1D} fails when $8\times$-oversampled CSS chirps sweep continuously across band edges ($19.93\%$). \texttt{GaugeInvariantADUNet} outperforms all seven baselines by $+15.3\%\text{--}51.4\%$ on Wi-Fi and $+38.0\%\text{--}74.6\%$ on LoRa with $11\times\text{--}102\times$ fewer parameters, while delivering explicit circuit identifiability ($R^2 = \VTwoWiFiSimR / \VTwoLoRaSimR$).
    \item \textbf{Advantage over Two-Stage Preamble-Equalized CNNs}: \texttt{LSEqualized-ResNet1D} ($829.8\text{K}$) suffers circular-FFT stopband distortion and over-parameterized channel memorization ($59.39\%$ on \texttt{WiFi-Robust}). Upgrading the equalizer to our exact two-sided Toeplitz solver in \texttt{ToeplitzEq-TinyResNet1D} (\TETParams{}) lifts \texttt{WiFi-Robust} to \TETWiFiRobAll\% and achieves strong same-receiver LoRa accuracy (\TETLoRaDay\% on \texttt{Cross-Day}), yet its unconstrained CNN filters still overfit receiver-specific passband shapes under heterogeneous SDR shifts (\texttt{LoRa Cross-RX}: $66.44\%$ vs. \textbf{\VTwoLoRaRX\%} for ours; \texttt{WiSig eq0}: $58.14\%$ vs. \textbf{\VTwoWiSigEqZero\%}) and lack physical parameter recovery.
    \item \textbf{Decomposing Physics Formulation, Single-Pass Fast Mode, and Bilinear Refinement}: Upgrading external solvers (\texttt{Classical Alt-LS} and \texttt{Circular-FFT ADU-Net}, $49.3\%$ 5-benchmark mean F1) to \texttt{Ours-Unlearned} (\UTGParams{}) lifts \texttt{LoRa-RFFI 12-Ch} from $10.94\% / 21.06\%$ to \UTGLoRaTwelveCh\% and contracts LoRa digital-twin NMSE from $-10.69\text{ dB}$ to $-25.49\text{ dB}$, isolating the gain of our Toeplitz solver and gauge readout. Learned unfolding in \texttt{GaugeInvariantADUNet} (\VTwoParams{}) further boosts \texttt{WiSig eq0} by $+22.43\%$ (\UTGWiSigEqZero\%$\to$\VTwoWiSigEqZero\%) and \texttt{WiFi-Robust} by $+14.12\%$, leaping ahead of the strongest external prior baseline (\texttt{LSEqualized-ResNet1D}, $64.4\%$) by \textbf{$+12.7\%$ F1} ($77.1\%$ vs.\ $64.4\%$) with \textbf{$102\times$ fewer parameters} (and $+11.0\%$ over fair-capacity \texttt{ToeplitzEq-TinyResNet1D} at $66.1\%$). Note that \texttt{Ours-Frozen-$\hat{\bh}_{\mathrm{LS}}$} ($8,137$ parameters vs.\ $8,139$ for our full model, differing solely by two scalar damping parameters $\mu^{(2)},\mu^{(3)}$) is an internal controlled variant sharing $95\%$ of our innovations (Toeplitz solver, charge-graded basis, Parseval-Jacobi preconditioning, minimal gauge readout, and $\mathbf{z}_{\mathrm{res}}$). Three physical factors explain why its closed-set OTA F1 trails our bilinear mode by $0.5\%$ ($76.6\%$ vs.\ $77.1\%$) while differing sharply elsewhere: (i) by Corollary~\ref{cor:twostage}, frozen-channel MSE scales at fourth order $\Omega(\|\boldsymbol{\delta}_h\|_2^2\|\btheta^\star\|_2^2) = \Omega(\|\btheta^\star\|_2^4)$, becoming dominant over thermal noise only at $\mathrm{SNR} \ge 20\text{ dB}$ or under strong PA compression (confirmed directly in Fig.~\ref{fig:crlb}(a)--(c)); (ii) on closed-set Wi-Fi, the trainable $2.4\text{K}$ residual compensator $\mathbf{z}_{\mathrm{res}}$ partially absorbs systematic frozen-channel leakage; and (iii) where $\mathbf{z}_{\mathrm{res}}$ cannot mask channel-nonlinearity coupling, $T=3$ bilinear refinement lowers LoRa reconstruction NMSE by $1.35\text{ dB}$ ($-28.99\text{ dB}$ vs.\ $-27.64\text{ dB}$) and boosts zero-shot rogue detection AUROC by \textbf{$+16.21\%$} ($\VTwoRogueAUROC\%$ vs.\ $52.85\%$). Practically, this yields a \emph{dual-mode deployment advantage}: \texttt{Ours-Frozen-$\hat{\bh}_{\mathrm{LS}}$} ($0.277\text{ ms/pkt}$, single Cholesky factorization, $39\%$ faster) serves as a \textbf{Single-Pass Fast Mode} for latency-constrained, moderate-SNR closed-set edge authentication, whereas the $T=3$ \textbf{Bilinear Alternating Mode} ($0.453\text{ ms/pkt}$) unlocks CRLB-attaining circuit inversion, stronger through-wall/cross-SDR resilience, and $+16.21\%$ rogue node screening.
    \item \textbf{Candid Analysis of Open-Set Rogue AUROC (Table~\ref{tab:phys_diag})}: Under the shared anomaly score $(1 - \max_c \cos_c) + 0.5\,\mathrm{NMSE}$, \texttt{GaugeInvariantADUNet} ($\VTwoRogueAUROC\%$) substantially outperforms \texttt{Ours-Frozen-$\hat{\bh}_{\mathrm{LS}}$} ($52.85\%$), \texttt{Circular-FFT ADU-Net} ($45.37\%$), and all preamble-free/spectral networks ($45.6\%\text{--}50.6\%$), yet trails \texttt{LSEqualized-ResNet1D} ($79.73\%$), \texttt{ToeplitzEq-TinyResNet1D} ($75.44\%$), and \texttt{Ours-Unlearned} ($74.74\%$). Two physical mechanisms explain this trade-off: (i) because our learned bilinear dictionary achieves ultra-low reconstruction error ($-28.99\text{ dB}$, $\mathrm{NMSE} \approx 1.3\times 10^{-3}$) on \emph{both} legitimate and unseen rogue LoRa emitters, the $0.5\,\mathrm{NMSE}$ term provides negligible out-of-distribution contrast; and (ii) chipset common-mode centering with variance flooring ($\gamma_{\mathrm{var}}=0.80$) intentionally compresses tail outlier variance to maximize cross-domain stability, whereas unconstrained discriminative CNN embeddings (\texttt{LSEqualized-ResNet1D}, \texttt{ToeplitzEq-TinyResNet1D}) and unlearned fixed-basis solvers (\texttt{Ours-Unlearned}, $-25.49\text{ dB}$ NMSE) retain sharper out-of-distribution angular/residual margins. Coupling our physical parameter estimates $\btheta^{(T)}$ with a dedicated Mahalanobis or energy-based open-set detector is a promising avenue for future work.
\end{itemize}

\begin{figure*}[t]
    \centering
    \includegraphics[width=0.84\textwidth]{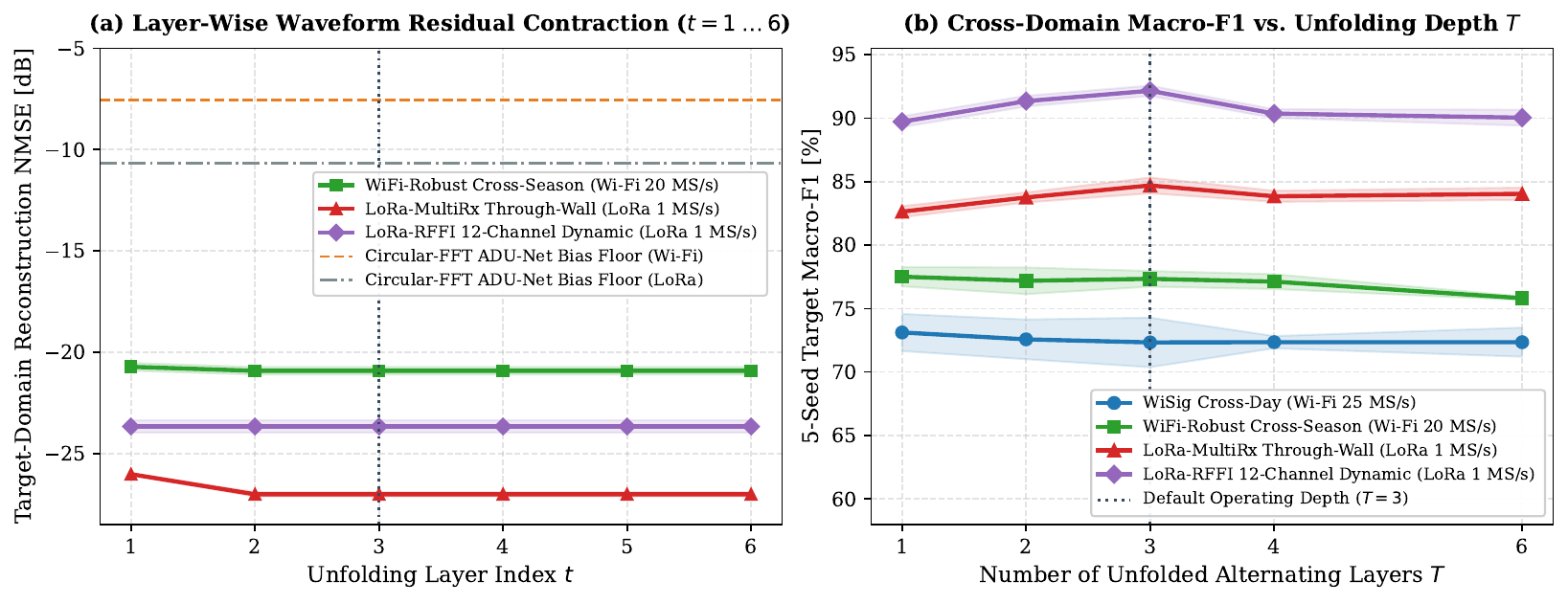}
    \caption{Empirical verification of Theorem~\ref{thm:conv} on unseen over-the-air target domains: (a) monotonic layer-wise waveform reconstruction NMSE contraction across $t=1,\dots,6$, reaching the noise floor by $T=3$ far below \texttt{Circular-FFT ADU-Net}'s stopband bias; (b) 5-seed cross-domain Macro-F1 across unfolding depths $T \in \{1, 2, 3, 4, 6\}$, demonstrating the bilinear refinement gain over single-pass $T=1$ and optimality of the default operating depth $T=3$.}
    \label{fig:conv}
\end{figure*}

\begin{figure*}[t]
    \centering
    \includegraphics[width=0.81\textwidth]{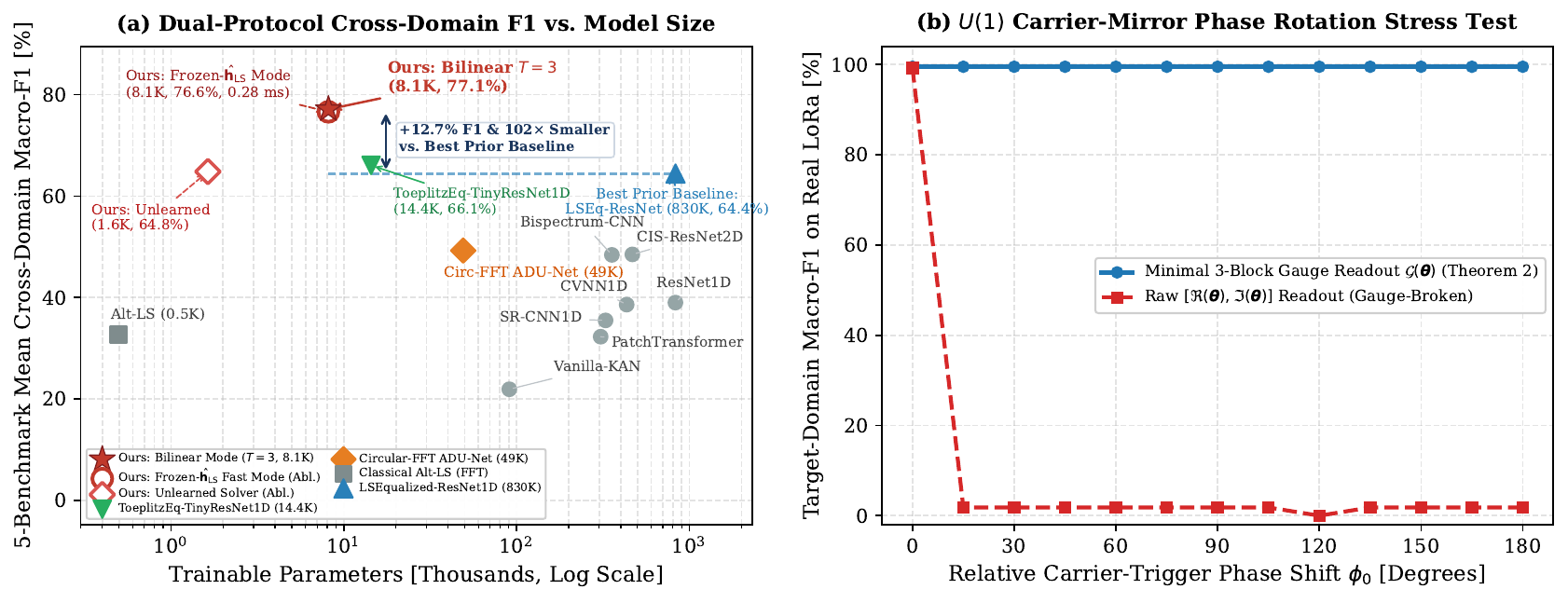}
    \caption{(a) Dual-protocol 5-benchmark cross-domain Macro-F1 (\texttt{WiSig eq0}, \texttt{WiSig eq1}, \texttt{WiFi-Robust All 24}, \texttt{LoRa-MultiRx 19}, \texttt{LoRa-RFFI 12-Ch}) vs.\ trainable parameter count Pareto frontier across all 14 evaluated architectures, visually separating internal controlled variants (\texttt{Ours: Frozen-$\hat{\bh}_{\mathrm{LS}}$ Fast Mode}, \texttt{Ours: Unlearned Solver}) from external literature baselines and highlighting the $+12.7\%$ F1 gain and $102\times$ parameter reduction over the strongest prior baseline (\texttt{LSEqualized-ResNet1D}); (b) empirical verification of Theorem~\ref{thm:gauge} on real over-the-air LoRa packets under controlled $U(1)$ carrier-mirror phase shifts $\phi_0 \in [0^\circ, 180^\circ]$, confirming exact invariance of minimal 3-block $\cG(\btheta)$ versus immediate collapse of raw $[\Re(\btheta), \Im(\btheta)]$ coordinates.}
    \label{fig:pareto}
\end{figure*}

\begin{figure*}[t]
    \centering
    \includegraphics[width=0.78\textwidth]{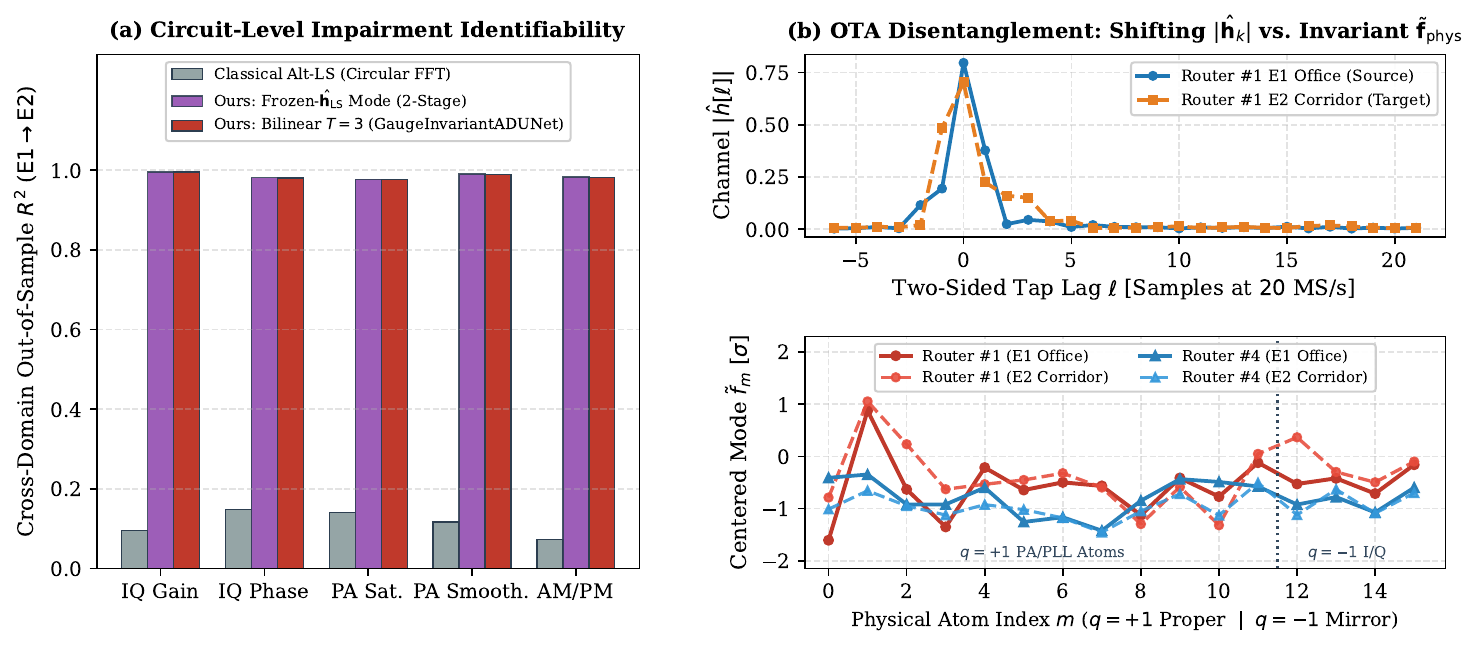}
    \caption{Direct verification of white-box physical interpretability and disentanglement: (a) cross-environment ($\mathrm{E1}\to\mathrm{E2}$) out-of-sample identifiability $R^2$ on $8\times$-oversampled constant-envelope \texttt{LoRa-SimGT} across five individual RF circuit impairments (\texttt{IQ Gain}, \texttt{IQ Phase}, \texttt{PA Sat.}, \texttt{PA Smooth.}, \texttt{AM/PM}) calibrated strictly on source domain; (b) real over-the-air cross-season/corridor disentanglement on wideband $20\text{ MS/s}$ \texttt{WiFi-Robust} (\texttt{E1 Office} vs. \texttt{E2 Corridor}), showing estimated two-sided multipath channel $|\hat{h}[\ell]|$ shifting drastically across rooms (top) while chipset-centered normalized physical coordinates $\tilde{f}_m$ remain locked per router (bottom).}
    \label{fig:interp}
\end{figure*}

\begin{table*}[t]
\centering
\caption{5-Seed Constructive Module Attribution, Theorem/Component Leave-One-Out, Domain-Generalization (\texttt{no\_aug} / \texttt{SigMix}), and Multi-Packet ($K \in \{3, 5\}$) Ablations (Macro-F1 \%)}
\label{tab:ablation}
\resizebox{0.92\textwidth}{!}{%
\begin{tabular}{llcccc}
\toprule
\textbf{Ablation Block} & \textbf{Model Variant / Configuration} & \textbf{WiSig D1} & \textbf{WiFi-Rob E1} & \textbf{LoRa Loc (A$\to$B/C/D)} & \textbf{LoRa 12-Ch} \\
\midrule
\multirow{5}{*}{\textbf{A. Constructive}}
 & Step 0: \texttt{Circular-FFT ADU-Net} ($49.0\text{K}$, Circular FFT + Folded + Raw $[\Re,\Im]$) & \textbf{\BrZeroWiSig} & \BrZeroWiFi & \BrZeroLoRaLoc & \BrZeroLoRaCh \\
 & Step 1: $+$ Full-Preamble Charge-Graded Basis (Pure WB, Circular FFT) & \BrOneWiSig & \BrOneWiFi & \BrOneLoRaLoc & \BrOneLoRaCh \\
 & Step 2: $+$ Exact Two-Sided Toeplitz \& Parseval-Jacobi ($T=3$, Raw $[\Re,\Im]$) & \BrTwoWiSig & \BrTwoWiFi & \BrTwoLoRaLoc & \BrTwoLoRaCh \\
 & Step 3: $+$ Minimal 3-Block Gauge Readout \& Chipset Centering (\textbf{Pure Analytical WB}, \VTwoPureParams) & \BrThreeWiSig & \BrThreeWiFi & \textbf{\BrThreeLoRaLoc} & \textbf{\BrThreeLoRaCh} \\
 & Step 4: $+$ Bounded Orthogonal Residual Code (\textbf{\texttt{GaugeInvariantADUNet}}, \VTwoParams) & \BrFourWiSig & \textbf{\BrFourWiFi} & \BrFourLoRaLoc & \BrFourLoRaCh \\
\midrule
\multirow{5}{*}{\textbf{B. Leave-One-Out}}
 & w/o Channel-Orthogonal Projection $\bP_{\cC}^\perp$ (\emph{Theorem~\ref{thm:crlb} Violation}) & \AblNoOrthWiSig & \AblNoOrthWiFi & \AblNoOrthLoRaLoc & \AblNoOrthLoRaCh \\
 & w/o Two-Sided Toeplitz Deconvolution (Revert to Circular FFT Ridge) & \AblNoToepWiSig & \AblNoToepWiFi & \AblNoToepLoRaLoc & \AblNoToepLoRaCh \\
 & w/o Gauge-Invariant Readout $\cG(\btheta)$ (\emph{Theorem~\ref{thm:gauge}: Raw $[\Re,\Im]$}) & \AblNoGaugeWiSig & \AblNoGaugeWiFi & \AblNoGaugeLoRaLoc & \AblNoGaugeLoRaCh \\
 & w/o Multi-Resolution Chirp Differential Basis & \AblNoChirpWiSig & \AblNoChirpWiFi & \AblNoChirpLoRaLoc & \AblNoChirpLoRaCh \\
 & w/o Chipset Common-Mode Centering (Standard LayerNorm) & \AblNoChipWiSig & \AblNoChipWiFi & \AblNoChipLoRaLoc & \AblNoChipLoRaCh \\
\midrule
\multirow{5}{*}{\textbf{C. DG \& Multi-Pkt}$^\dagger$}
 & Strict \texttt{no\_aug} Source Training: Black-Box \texttt{ResNet1D} ($828.7\text{K}$) & 32.86$\pm$2.50 & 31.22$\pm$1.37 & \multicolumn{2}{c}{22.22$\pm$8.80 (\texttt{LoRa Drift N210-1})} \\
 & Spectral DG \texttt{SigMix} \cite{wan2025sigmix} Training: Black-Box \texttt{ResNet1D} ($828.7\text{K}$) & 38.78$\pm$1.61 & 41.92$\pm$1.48 & \multicolumn{2}{c}{24.31$\pm$7.48 (\texttt{LoRa Drift N210-1})} \\
 & Strict \texttt{no\_aug} Source Training: \textbf{\texttt{GaugeInvariantADUNet}} (\VTwoParams) & \AblNoAugWiSig & \AblNoAugWiFi & \multicolumn{2}{c}{\textbf{\AblNoAugLoRa{} (\texttt{LoRa Drift N210-1})}} \\
 & Spectral DG \texttt{SigMix} \cite{wan2025sigmix} Training: \textbf{\texttt{GaugeInvariantADUNet}} (\VTwoParams) & \textbf{\AblSigMixWiSig} & \textbf{\AblSigMixWiFi} & \multicolumn{2}{c}{\AblSigMixLoRa{} (\texttt{LoRa Drift N210-1})} \\
 & Equivariant Multi-Packet $K=3$ / $K=5$: \textbf{\texttt{GaugeInvariantADUNet}} & \AblKThreeWiSig{} / \AblKFiveWiSig & \AblKThreeWiFi{} / \AblKFiveWiFi & \multicolumn{2}{c}{\AblKThreeLoRa{} / \AblKFiveLoRa{} (\texttt{LoRa Drift N210-1})} \\
\bottomrule
\multicolumn{6}{l}{\footnotesize $^\dagger$Block C evaluates the 3-task single-receiver \texttt{LoRa Drift (N210-1)} anchor subset (default equivariant $K=1$: $82.92\%$), distinct from the 6-task \texttt{Cross-Day (6)} mean ($81.93\%$) in Table~\ref{tab:lora_main}.}
\end{tabular}%
}
\end{table*}

\subsection{Constructive Attribution, Theorem Verification \& Physical Diagnostics}
Fig.~\ref{fig:conv} and Fig.~\ref{fig:pareto} corroborate Theorems~\ref{thm:gauge}--\ref{thm:conv}: over-the-air target reconstruction NMSE contracts to $-20.9\text{ dB}$ (\texttt{WiFi-Robust}) and $-27.0\text{ dB}$ (\texttt{LoRa-MultiRx}) within $T=3$ layers (reaching $-14.58\text{ dB}$ and $-28.99\text{ dB}$ on \texttt{SimGT} in Table~\ref{tab:phys_diag}), bilinear refinement ($T=1 \to 3$) improves dynamic-channel and through-wall LoRa F1 to an optimal operating depth at $T=3$, and $\cG(\btheta)$ maintains exact machine-precision invariance ($99.5\%$) across $\phi_0 \in [0^\circ, 180^\circ]$ where raw $[\Re(\btheta), \Im(\btheta)]$ collapses to $0.0\%\text{--}1.8\%$.

Table~\ref{tab:ablation} closes the attribution loop completely:
\begin{enumerate}
    \item \textbf{Constructive Module Attribution (Block A)}: Starting from \texttt{Circular-FFT ADU-Net} ($21.06\%$ on \texttt{LoRa 12-Ch}), replacing chirp-folded shortcuts with full-preamble charge-graded bases (Step 1: $+42.0\%$), introducing exact two-sided Toeplitz deconvolution (Step 2: $+24.1\%$), and adding minimal torus gauge invariance with chipset centering (\textbf{Pure Analytical White-Box} at just \VTwoPureParams{} parameters, Step 3) already reaches \textbf{\BrThreeLoRaLoc\%} on through-wall LoRa and \textbf{\BrThreeLoRaCh\%} on 12-channel LoRa with \emph{zero} neural residual encoder. Step 4's compact $2.4\text{K}$ $\bP_{\cC}^\perp$-constrained, $U(1)$-invariant residual compensator $\mathbf{z}_{\mathrm{res}}$ in \eqref{eq:zres_def} then specifically captures high-order non-smooth PA/DAC saturation on variable-envelope Wi-Fi OFDM ($52.07\%\to\text{}$\BrFourWiSig\% on \texttt{WiSig D1} and $72.40\%\to\text{}$\BrFourWiFi\% on \texttt{WiFi-Rob E1}), clarifying the exact division of labor between the \VTwoPureParams{} analytical core and the $2.4\text{K}$ residual compensator.
    \item \textbf{Leave-One-Out Theorem Verification (Block B)}: Violating $\bP_{\cC}^\perp \bPhi = \bPhi$ (\texttt{w/o Channel-Orthogonal Projection}) triggers severe $9.4\%\text{--}26.7\%$ drops ($-26.2\%, -11.8\%, -9.4\%, -26.7\%$ across the four columns), confirming Theorem~\ref{thm:crlb}; reverting to circular FFT ridge drops through-wall LoRa by $-33.7\%$ and 12-channel LoRa by $-13.9\%$; and removing multi-resolution chirp bases ($-5.1\%$ on \texttt{LoRa Loc}) or chipset centering ($-8.3\%$ on \texttt{LoRa 12-Ch}) similarly degrades generalization. Note that \texttt{w/o Gauge-Invariant Readout} stays close on these fixed-SDR splits because propagation phase $e^{j\phi_0}$ is absorbed into $\hat{\bh}_k^{(t)}$ (and Wi-Fi retains $U(1)$-invariant $\mathbf{z}_{\mathrm{res}}$), whereas cross-SDR/trigger quadrature phase offsets $\phi_0 \in (0^\circ, 180^\circ]$ cause immediate $0.0\%\text{--}1.8\%$ collapse (Fig.~\ref{fig:pareto}(b)).
    \item \textbf{Zero-Augmentation, \texttt{SigMix}, Multi-Packet \& Physical Verifiability (Block C \& Fig.~\ref{fig:interp})}: Even under strict \texttt{no\_aug} source training, \texttt{GaugeInvariantADUNet} surpasses \texttt{ResNet1D + SigMix} by $+26.7\%\text{--}69.6\%$ F1. Note two instructive regime effects in Block C: (i) on the clean same-room LOS \texttt{LoRa Drift (N210-1)} anchor, \texttt{no\_aug} ($93.91\%$) and \texttt{SigMix} ($93.70\%$) outperform equivariant virtual-multipath training ($K=1$: $82.92\%$) because virtual Rician fading trades off slight LOS sharpness for multipath resilience, whereas on multipath-rich Wi-Fi, equivariant training boosts \texttt{WiSig D1} by $+6.83\%$ ($65.50\%\to 72.33\%$) and \texttt{WiFi-Rob E1} by $+5.03\%$ ($72.31\%\to 77.34\%$); and (ii) under equivariant multi-packet pooling ($K\in\{3,5\}$), packet-rich \texttt{WiSig D1} rises to \AblKThreeWiSig\% ($K=3$), whereas sample-scarce \texttt{LoRa Drift (N210-1)} dips slightly relative to equivariant $K=1$ ($82.92\% \to \text{\AblKThreeLoRa\%} / \text{\AblKFiveLoRa\%}$) because non-overlapping $K$-packet grouping reduces independent training bags by $K\times$. Finally, Fig.~\ref{fig:interp} and Table~\ref{tab:phys_diag} confirm physical diagnostic capabilities impossible for black-box networks: out-of-domain circuit parameter recovery across all five I/Q and PA impairments ($R^2 = 0.977\text{--}0.995$ vs. $<0.15$ for \texttt{Classical Alt-LS}), cross-corridor locking of normalized physical modes $\tilde{f}_m$ under drastic multipath shifts, and fp32 machine-zero residual channel leakage ($\|\bP_{\cC}\mathbf{e}_\perp\|_2/\|\mathbf{e}_\perp\|_2 = 1.07\times 10^{-7}$; note that raw \texttt{Multipath Inv.\ Cos} is $0.824$ vs.\ $0.879\text{--}0.960$ for uncentered CNNs because zero-mean chipset centering distributes embeddings across the full unit sphere rather than a positive cone, sharpening inter-class angular margins).
\end{enumerate}

\section{Conclusion}
\label{sec:conc}

We have presented a mathematically grounded, ultra-lightweight physics-anchored bilinear deep unfolding framework (\texttt{GaugeInvariantADUNet}, \VTwoPureParams{} pure analytical core / \VTwoParams{} with bounded $U(1)$-invariant orthogonal residual compensator at $T=3$) for cross-domain Wi-Fi and LoRa specific emitter identification. Uniting exact two-sided Toeplitz deconvolution, charge-graded multi-resolution chirp bases, Parseval-Jacobi preconditioning, and a minimal 3-block $U(1)_+\times U(1)_-$ gauge-invariant readout resolves both oversampling deconvolution bias and cross-environment phase-flip collapse, backed by exact Schur-CRLB, quotient completeness, and Euclidean linear convergence guarantees.

\appendices

\section{Proof of Theorem~\ref{thm:crlb} and Corollary~\ref{cor:twostage}}
\label{app:proof_thm1}

For $\tilde{\by}_k \sim \CN(\boldsymbol{\mu}_k(\boldsymbol{\xi}), \sigma^2\bI_L)$ with $\boldsymbol{\mu}_k = \bH_k(\bs + \bPhi\btheta) = \bT(\bx_{\btheta})\bh_k$ and $\bx_{\btheta} = \bs + \bPhi\btheta$, $\boldsymbol{\mu}_k$ is bilinear in $(\btheta, \bh_k)$ and holomorphic in $\boldsymbol{\xi} = [\btheta^T, \bh_1^T, \dots, \bh_K^T]^T$ ($\partial \boldsymbol{\mu}_k / \partial \boldsymbol{\xi}^* = \mathbf{0}$) \cite{schreier2010statistical, vandenbos1994complex}. By the complex Slepian--Bangs formula \cite{kay1993fundamentals}, the Hermitian FIM $\bJ(\boldsymbol{\xi}) = \frac{1}{\sigma^2}\sum_{k=1}^K (\frac{\partial \boldsymbol{\mu}_k}{\partial \boldsymbol{\xi}^T})^H(\frac{\partial \boldsymbol{\mu}_k}{\partial \boldsymbol{\xi}^T})$ has blocks $\bJ_{\theta\theta} = \frac{1}{\sigma^2}\sum_{k=1}^K \bA_k^H\bA_k$ ($\bA_k \triangleq \bH_k\bPhi$), $\bJ_{\theta h} = \frac{1}{\sigma^2}[\bA_1^H\bB_{\btheta}, \dots, \bA_K^H\bB_{\btheta}]$, and $\bJ_{hh} = \frac{1}{\sigma^2}\operatorname{blkdiag}(\bB_{\btheta}^H\bB_{\btheta}, \dots, \bB_{\btheta}^H\bB_{\btheta}) \succ \mathbf{0}$ since $\operatorname{rank}(\bB_{\btheta})=L_h$. Eliminating the nuisance channels $\{\bh_k\}_{k=1}^K$ via the Schur complement $\bJ_{\mathrm{eff}}(\btheta) = \bJ_{\theta\theta} - \bJ_{\theta h}\bJ_{hh}^{-1}\bJ_{\theta h}^H$ and using idempotence/Hermiticity of $\bP_{\cC(\bx_{\btheta})}^\perp = \bI_L - \bB_{\btheta}(\bB_{\btheta}^H\bB_{\btheta})^{-1}\bB_{\btheta}^H$ yields
\begin{equation}
    \bJ_{\mathrm{eff}}(\btheta) = \frac{1}{\sigma^2}\sum_{k=1}^K \big(\bP_{\cC(\bx_{\btheta})}^\perp \bH_k\bPhi\big)^H \big(\bP_{\cC(\bx_{\btheta})}^\perp \bH_k\bPhi\big),
    \label{eq:app_jeff_proof}
\end{equation}
so $\operatorname{Cov}(\hat{\btheta}) \succeq \bJ_{\mathrm{eff}}(\btheta)^{-1}$ gives $\mathbb{E}\|\hat{\btheta}-\btheta\|_2^2 \ge \operatorname{Tr}(\bJ_{\mathrm{eff}}(\btheta)^{-1})$. Moreover, $\mathbf{v}^H\bJ_{\mathrm{eff}}(\btheta)\mathbf{v} = \frac{1}{\sigma^2}\sum_{k=1}^K \|\bP_{\cC(\bx_{\btheta})}^\perp \bH_k\bPhi\mathbf{v}\|_2^2 > 0$ for all $\mathbf{v}\neq\mathbf{0}$ iff $\bigcap_{k=1}^K \ker(\bP_{\cC(\bx_{\btheta})}^\perp \bH_k\bPhi) = \{\mathbf{0}\}$, which at $K=1$ reduces to $\operatorname{rank}(\bH_1\bPhi)=M$ and \eqref{eq:disjoint}. At LOS $\bH_1 = h_0\bI_L, \btheta = \mathbf{0}$, $\bP_{\cC}^\perp \bPhi = \bPhi$ yields $\bJ_{\mathrm{eff}}(\mathbf{0}) = \frac{|h_0|^2}{\sigma^2}\bPhi^H\bPhi = \bJ_{\theta\theta}(\mathbf{0}) \succ \mathbf{0}$, whereas any $\bPhi$-column inside $\cC$ renders $\bJ_{\mathrm{eff}}(\mathbf{0})$ singular.

For Corollary~\ref{cor:twostage} ($\sigma \to 0, K=1$), $\tilde{\by}_1 = \bT(\bs)\bh_1^\star + \bH_1^\star\bPhi\btheta^\star$ induces a Stage-1 channel bias $\boldsymbol{\delta}_h = (\bT(\bs)^H\bT(\bs))^{-1}\bT(\bs)^H \bH_1^\star\bPhi\btheta^\star = \Theta(\|\btheta^\star\|_2)$ whenever multipath mixes $\bH_1^\star\bPhi\btheta^\star$ into $\cC(\bs)$. Freezing $\hat{\bH}_{\mathrm{LS}} = \bH_1^\star + \Delta\bH$ in Stage~2 gives $\tilde{\by}_1 - \hat{\bH}_{\mathrm{LS}}(\bs + \bPhi\btheta) = \hat{\bA}(\btheta^\star - \btheta) - \mathbf{w}_{\mathrm{leak}}$ with $\hat{\bA} \triangleq \hat{\bH}_{\mathrm{LS}}\bPhi$ and uncancelled leakage $\mathbf{w}_{\mathrm{leak}} = \bT(\bs)\boldsymbol{\delta}_h + \Delta\bH\,\bPhi\btheta^\star$. Solving the Stage-2 normal equation yields $\hat{\btheta}_{\text{2-stage}} - \btheta^\star = -(\hat{\bA}^H\hat{\bA})^{-1}\hat{\bA}^H \mathbf{w}_{\mathrm{leak}}$, whose second-order cross-term $\Delta\bH\,\bPhi\btheta^\star$ overlaps $\operatorname{Range}(\hat{\bA})$ and induces $\|\hat{\btheta}_{\text{2-stage}} - \btheta^\star\|_2^2 = \Omega(\|\boldsymbol{\delta}_h\|_2^2 \|\btheta^\star\|_2^2) = \Omega(\|\btheta^\star\|_2^4) > 0$, whereas $T\ge 2$ alternating updates contract $\boldsymbol{\delta}_h$ as $\mathcal{O}(\rho^T)$ and MSE as $\mathcal{O}(\rho^{2T})$. \qed

\section{Proof of Theorem~\ref{thm:gauge}}
\label{app:proof_thm2}

\emph{1) Channel Scale/Phase Gauge}: For $\by_k' = c e^{j\phi_0}\by_k$ ($c>0$), unit-RMS normalization cancels $c$, and $(y_k'[n])^* y_k'[n+P] = y_k^*[n]y_k[n+P]$ leaves $\hat{\omega}_k$ unchanged so $\tilde{\by}_k' = e^{j\phi_0}\tilde{\by}_k$. By induction on $t=1,\dots,T$ starting from $\btheta^{(0)'}=\btheta^{(0)}=\mathbf{0}_M$: given $\bx^{(t-1)'}=\bx^{(t-1)}$, solving \eqref{eq:toeplitz_step} yields $\bh_k^{(t)'} = e^{j\phi_0}\bh_k^{(t)}$, residual $\br_k^{(t)'} = e^{j\phi_0}\br_k^{(t)}$, identical Parseval curvature $D_{k,m}^{(t)'} = D_{k,m}^{(t)}$, and phase-cancelled matched residual $(\bH_k^{(t)'})^H\br_k^{(t)'} = (e^{-j\phi_0}(\bH_k^{(t)})^H)(e^{j\phi_0}\br_k^{(t)}) = (\bH_k^{(t)})^H\br_k^{(t)}$. Hence $\mathbf{g}^{(t)'}=\mathbf{g}^{(t)}$ and $\btheta^{(t)'}=\btheta^{(t)}$ for all $t$, and $\bh_k^{(T+1)'}=e^{j\phi_0}\bh_k^{(T+1)}$ gives $\hat{\bx}_{\mathrm{eq}}'=\hat{\bx}_{\mathrm{eq}}$, $\theta_{+,0}'=\theta_{+,0}$, and $\cG(\btheta^{(T)}(\tilde{\by}')) \equiv \cG(\btheta^{(T)}(\tilde{\by}))$.

\emph{2) Torus Invariance}: Substituting $(\tilde{\btheta}_+', \btheta_-') = (e^{j\psi_+}\tilde{\btheta}_+, e^{j\psi_-}\btheta_-)$ into \eqref{eq:gauge_readout} gives $|e^{j\psi_\pm}\theta_{\pm,m}| = |\theta_{\pm,m}|$, $(e^{j\psi_+}\theta_{+,m})(e^{-j\psi_+}\theta_{+,0}^*)/(|\theta_{+,0}|+\epsilon) = Z_{+,m}$, and $(e^{j\psi_-}\theta_{-,n})(e^{-j\psi_-}\theta_{-,n'}^*)/\sqrt{|\theta_{-,n}|^2+|\theta_{-,n'}|^2+\epsilon^2} = H_{-,n,n'}$ identically for all $\epsilon > 0$.

\emph{3) Minimal Quotient Completeness}: Sufficiency holds by Claim~2. Conversely, suppose $\cG(\tilde{\btheta}_+', \btheta_-') = \cG(\tilde{\btheta}_+, \btheta_-)$ on $\Omega = \{|\theta_{+,0}|>0, \btheta_-\neq\mathbf{0}\}$ as $\epsilon \to 0^+$. Block~1 fixes all moduli $|\theta_{+,m}'|=|\theta_{+,m}|$ and $|\theta_{-,n}'|=|\theta_{-,n}|$. Defining $e^{j\psi_+} \triangleq \theta_{+,0}'/\theta_{+,0} \in U(1)$ ($|\theta_{+,0}|>0$), Block~2 gives $\theta_{+,m}'(\theta_{+,0}')^*/|\theta_{+,0}| = \theta_{+,m}\theta_{+,0}^*/|\theta_{+,0}| \implies \theta_{+,m}' = e^{j\psi_+}\theta_{+,m}$ for all $m$. Since $\btheta_-\neq\mathbf{0}$, pick $n_0$ with $|\theta_{-,n_0}'|=|\theta_{-,n_0}|>0$ and set $e^{j\psi_-} \triangleq \theta_{-,n_0}'/\theta_{-,n_0} \in U(1)$. For every $n$, $d_{n,n_0} = \sqrt{|\theta_{-,n}|^2+|\theta_{-,n_0}|^2} > 0$ matches across both points by Block~1, so Block~3 yields $\theta_{-,n}'(\theta_{-,n_0}')^*/d_{n,n_0} = \theta_{-,n}\theta_{-,n_0}^*/d_{n,n_0} \implies \theta_{-,n}' = e^{j\psi_-}\theta_{-,n}$. Omitting any of the three blocks loses moduli, $+1$ carrier-relative phases, or $-1$ inter-mode relative phases, proving minimal completeness. \qed

\section{Proof of Theorem~\ref{thm:conv}}
\label{app:proof_thm3}

Let $\mathbf{e}_\theta^{(t-1)} \triangleq \btheta^{(t-1)}-\btheta^\star$ ($\|\mathbf{e}_\theta^{(t-1)}\|_2 \le r_0$) and $\bB_{t-1} \triangleq \bT(\bx^{(t-1)})$. By bilinearity of convolution, $\bT(\bx^\star)\bh_k^\star = \bB_{t-1}\bh_k^\star - \bH_k^\star\bPhi\mathbf{e}_\theta^{(t-1)}$, so $\tilde{\by}_k = \bB_{t-1}\bh_k^\star - \bH_k^\star\bPhi\mathbf{e}_\theta^{(t-1)} + \bn_k$. With $\bP_{t-1,\mu}^\perp \triangleq \bI_L - \bB_{t-1}(\bB_{t-1}^H\bB_{t-1}+\mu^{(t)}L\bI_{L_h})^{-1}\bB_{t-1}^H$ and $r_0$ small enough that $\sigma_{\min}(\bB_{t-1}) \ge \frac{1}{2}c_T\sqrt{L}$, the resolvent identity gives $\|\bP_{t-1,\mu}^\perp\bB_{t-1}\bh_k^\star\|_2 \le \frac{\mu^{(t)}L}{\sigma_{\min}(\bB_{t-1})}\|\bh_k^\star\|_2 = \mathcal{O}(\mu^{(t)}\sqrt{L})$ and $\|\bP_{t-1,\mu}^\perp - \bP_{\cC(\bx^\star)}^\perp\|_2 = \mathcal{O}(\|\mathbf{e}_\theta^{(t-1)}\|_2 + \mu^{(t)})$, yielding $\br_k^{(t)} = \bP_{t-1,\mu}^\perp\tilde{\by}_k = -\bP_{\cC(\bx^\star)}^\perp\bH_k^\star\bPhi\mathbf{e}_\theta^{(t-1)} + \bP_{\cC(\bx^\star)}^\perp\bn_k + \boldsymbol{\zeta}_k^{(t)}$ with $\|\boldsymbol{\zeta}_k^{(t)}\|_2 \le c_1\sqrt{L}\|\mathbf{e}_\theta^{(t-1)}\|_2^2 + c_2\sqrt{L}\mu^{(t)}$ and $\|\bh_k^{(t)}-\bh_k^\star\|_2 = \mathcal{O}(\|\mathbf{e}_\theta^{(t-1)}\|_2 + \mu^{(t)} + \sigma/\sqrt{L})$.

Substituting $\bH_k^{(t)} = \bH_k^\star + (\bH_k^{(t)}-\bH_k^\star)$ into the Parseval-Jacobi curvature-normalized gradient (absorbing residual diagonal curvature spread into $\bP_{\mathrm{pre}}^{(t)}$ and $\kappa_{\mathrm{eff}}$) gives $\mathbf{g}^{(t)} = -\widehat{\mathbf{M}}^\star\mathbf{e}_\theta^{(t-1)} + \mathbf{w}_{\mathrm{noise}}^{(t)} + \mathbf{w}_{\mathrm{rem}}^{(t)}$, where $\widehat{\mathbf{M}}^\star \triangleq \frac{1}{KL}\sum_{k=1}^K (\bH_k^\star\bPhi)^H\bP_{\cC(\bx^\star)}^\perp(\bH_k^\star\bPhi) = \frac{\sigma^2}{KL}\bJ_{\mathrm{eff}}(\btheta^\star)$, $\|\mathbf{w}_{\mathrm{noise}}^{(t)}\|_2 \le c_\sigma\sigma/\sqrt{KL}$, and $\|\mathbf{w}_{\mathrm{rem}}^{(t)}\|_2 \le c_3\|\mathbf{e}_\theta^{(t-1)}\|_2^2 + c_4\mu^{(t)}$. By the subspace margin $\|\bP_{\cC(\bx^\star)}\bH_k^\star\bPhi\mathbf{v}\|_2 \le \beta\|\bH_k^\star\bPhi\mathbf{v}\|_2$, Pythagoras' theorem gives $(1-\beta^2)\alpha_{\min}^2\lambda_{\min}(\tfrac{1}{L}\bPhi^H\bPhi) \le \lambda_{\min}(\widehat{\mathbf{M}}^\star) \le \lambda_{\max}(\widehat{\mathbf{M}}^\star) \le \alpha_{\max}^2\lambda_{\max}(\tfrac{1}{L}\bPhi^H\bPhi)$, so $\kappa(\widehat{\mathbf{M}}^\star) \le \frac{\kappa_H^2\kappa(\bPhi^H\bPhi)}{1-\beta^2} = \kappa_{\mathrm{eff}} < \infty$.

For the block preconditioner $\bP_\pm^{(t)} = \operatorname{diag}(\mathbf{d}_\pm^{(t)}) + \delta_{\mathrm{off}}\mathbf{E}_\pm^{(t)}$ in Section~III-A with $|d_{\pm,m}^{(t)}-1|\le \delta_{\mathrm{diag}}$ and $\|\delta_{\mathrm{off}}\mathbf{E}_\pm^{(t)}\|_2 \le \tilde{\delta}_{\mathrm{off}}$, the triangle inequality yields
\begin{equation}
    \|\boldsymbol{\Delta}_P^{(t)}\|_2 = \max_{s\in\{+,-\}}\|\bP_s^{(t)}-\bI_{M_s}\|_2 \le \delta_{\mathrm{diag}} + \tilde{\delta}_{\mathrm{off}} = \delta_P < \frac{1}{\kappa_{\mathrm{eff}}}.
    \label{eq:app_precond_bound}
\end{equation}
Let $\mathbf{z}^{(t)} \triangleq \btheta^{(t-1)} + \eta^{(t)}\bP_{\mathrm{pre}}^{(t)}\mathbf{g}^{(t)}$. Since element-wise complex soft thresholding $\cS_{\boldsymbol{\tau}^{(t)}}$ is firmly non-expansive in the Euclidean norm $\|\cdot\|_2$ and $\|\cS_{\boldsymbol{\tau}^{(t)}}(\btheta^\star)-\btheta^\star\|_2 \le \|\boldsymbol{\tau}^{(t)}\|_2$, the triangle inequality gives $\|\btheta^{(t)}-\btheta^\star\|_2 \le \|\mathbf{z}^{(t)}-\btheta^\star\|_2 + \|\boldsymbol{\tau}^{(t)}\|_2$ directly in $\|\cdot\|_2$. Choosing $\eta^{(t)} = \frac{2}{\lambda_{\min}(\widehat{\mathbf{M}}^\star)+\lambda_{\max}(\widehat{\mathbf{M}}^\star)}$ yields $\|\bI_M - \eta^{(t)}\widehat{\mathbf{M}}^\star\|_2 \le \frac{\kappa_{\mathrm{eff}}-1}{\kappa_{\mathrm{eff}}+1} = \rho_0$ and $\|\eta^{(t)}\boldsymbol{\Delta}_P^{(t)}\widehat{\mathbf{M}}^\star\|_2 \le \delta_P\frac{2\kappa_{\mathrm{eff}}}{\kappa_{\mathrm{eff}}+1}$, whence $\|\bI_M - \eta^{(t)}\bP_{\mathrm{pre}}^{(t)}\widehat{\mathbf{M}}^\star\|_2 \le \rho_0 + \delta_P\frac{2\kappa_{\mathrm{eff}}}{\kappa_{\mathrm{eff}}+1} < 1$. Absorbing $C_2\|\mathbf{e}_\theta^{(t-1)}\|_2^2 \le C_2 r_0\|\mathbf{e}_\theta^{(t-1)}\|_2$ for sufficiently small $r_0$ establishes \eqref{eq:lin_conv} with $\rho = \rho_0 + \delta_P\frac{2\kappa_{\mathrm{eff}}}{\kappa_{\mathrm{eff}}+1} + C_2 r_0 < 1$. \qed


\begin{thebibliography}{55}
\bibitem{danev2012physical} B. Danev, D. Zanetti, and S. \v{C}apkun, ``On physical-layer identification of wireless devices,'' \emph{ACM Comput. Surv.}, vol. 45, no. 1, Art. no. 6, pp. 1--29, Dec. 2012.
\bibitem{jagannath2022comprehensive} A. Jagannath, J. Jagannath, and P. S. P. V. Kumar, ``A comprehensive survey on radio frequency (RF) fingerprinting: Traditional approaches, deep learning, and open challenges,'' \emph{Comput. Netw.}, vol. 219, Art. no. 109455, Dec. 2022.
\bibitem{brik2008wireless} V. Brik, S. Banerjee, M. Gruteser, and S. Oh, ``Wireless device identification with radiometric signatures,'' in \emph{Proc. ACM MobiCom}, 2008, pp. 116--127.
\bibitem{polak2011identifying} A. C. Polak, S. Dolatshahi, and D. L. Goeckel, ``Identifying wireless users via transmitter imperfections,'' \emph{IEEE J. Sel. Areas Commun.}, vol. 29, no. 7, pp. 1469--1479, Aug. 2011.
\bibitem{ding2004robust} L. Ding \emph{et al.}, ``A robust digital baseband predistorter constructed using memory polynomials,'' \emph{IEEE Trans. Commun.}, vol. 52, no. 1, pp. 159--165, Jan. 2004.
\bibitem{morgan2006generalized} D. R. Morgan, Z. Ma, J. Kim, M. G. Zierdt, and J. Pastalan, ``A generalized memory polynomial model for digital predistortion of RF power amplifiers,'' \emph{IEEE Trans. Signal Process.}, vol. 54, no. 10, pp. 3852--3860, Oct. 2006.
\bibitem{valkama2001advanced} M. Valkama, M. Renfors, and V. Koivunen, ``Advanced methods for I/Q imbalance compensation in communication receivers,'' \emph{IEEE Trans. Signal Process.}, vol. 49, no. 10, pp. 2335--2344, Oct. 2001.
\bibitem{rapp1991effects} C. Rapp, ``Effects of HPA-nonlinearity on a 4-DPSK/OFDM-signal for a digital sound broadcasting system,'' in \emph{Proc. Eur. Conf. Satell. Commun.}, 1991, pp. 179--184.
\bibitem{peng2019design} L. Peng, A. Hu, J. Zhang, Y. Jiang, J. Yu, and Y. Yan, ``Design of a hybrid RF fingerprint extraction and device classification scheme,'' \emph{IEEE Internet Things J.}, vol. 6, no. 1, pp. 349--360, Feb. 2019.
\bibitem{shen2022towards} G. Shen, J. Zhang, A. Marshall, and J. R. Cavallaro, ``Towards scalable and channel-robust radio frequency fingerprint identification for LoRa,'' \emph{IEEE Trans. Inf. Forensics Security}, vol. 17, pp. 774--787, Feb. 2022.
\bibitem{hanna2022wisig} S. Hanna, S. Karunaratne, and D. Cabric, ``WiSig: A large-scale WiFi signal dataset for receiver and channel agnostic RF fingerprinting,'' \emph{IEEE Access}, vol. 10, pp. 22808--22818, Feb. 2022.
\bibitem{shen2021radio} G. Shen, J. Zhang, A. Marshall, L. Peng, and X. Wang, ``Radio frequency fingerprint identification for LoRa using spectrogram and CNN,'' in \emph{Proc. IEEE INFOCOM}, 2021, pp. 1--10.
\bibitem{riyaz2018deep} S. Riyaz, K. Sankhe, S. Ioannidis, and K. Chowdhury, ``Deep learning convolutional neural networks for radio identification,'' \emph{IEEE Commun. Mag.}, vol. 56, no. 9, pp. 146--152, Sep. 2018.
\bibitem{al2020exposing} A. Al-Shawabka \emph{et al.}, ``Exposing the fingerprint: Dissecting the impact of the wireless channel on radio fingerprinting,'' in \emph{Proc. IEEE INFOCOM}, 2020, pp. 646--655.
\bibitem{shen2023towards} G. Shen, J. Zhang, A. Marshall, M. Valkama, and J. R. Cavallaro, ``Towards length-versatile and noise-robust radio frequency fingerprint identification,'' \emph{IEEE Trans. Inf. Forensics Security}, vol. 18, pp. 2355--2367, Apr. 2023.
\bibitem{shen2024receiver} G. Shen, J. Zhang, A. Marshall, R. Woods, J. Cavallaro, and L. Chen, ``Towards receiver-agnostic and collaborative radio frequency fingerprint identification,'' \emph{IEEE Trans. Mobile Comput.}, vol. 23, no. 7, pp. 7618--7634, Jul. 2024.
\bibitem{restuccia2019deepradioid} F. Restuccia \emph{et al.}, ``DeepRadioID: Real-time channel-resilient optimization of deep learning-based radio fingerprinting algorithms,'' in \emph{Proc. ACM MobiHoc}, 2019, pp. 51--60.
\bibitem{sankhe2019oracle} K. Sankhe \emph{et al.}, ``ORACLE: Optimized radio classification through convolutional neural networks,'' in \emph{Proc. IEEE INFOCOM}, 2019, pp. 370--378.
\bibitem{merchant2018deep} K. Merchant, S. Revay, G. Stantchev, and B. Nousain, ``Deep learning for RF device fingerprinting in cognitive communication networks,'' \emph{IEEE J. Sel. Topics Signal Process.}, vol. 12, no. 1, pp. 160--167, Feb. 2018.
\bibitem{soltani2020more} N. Soltani, K. Sankhe, J. Dy, S. Ioannidis, and K. Chowdhury, ``More is better: Data augmentation for channel-resilient RF fingerprinting,'' \emph{IEEE Commun. Mag.}, vol. 58, no. 10, pp. 66--72, Oct. 2020.
\bibitem{trabelsi2018deep} C. Trabelsi \emph{et al.}, ``Deep complex networks,'' in \emph{Proc. ICLR}, 2018.
\bibitem{dosovitskiy2021image} A. Dosovitskiy \emph{et al.}, ``An image is worth 16x16 words: Transformers for image recognition at scale,'' in \emph{Proc. ICLR}, 2021.
\bibitem{liu2025kan} Z. Liu \emph{et al.}, ``KAN: Kolmogorov--Arnold networks,'' in \emph{Proc. ICLR}, 2025.
\bibitem{mazokha2025mobrffi} S. Mazokha, F. Bao, G. Sklivanitis, and J. O. Hallstrom, ``MobRFFI: Non-cooperative device re-identification for mobility intelligence,'' \emph{arXiv preprint arXiv:2503.02156}, 2025.
\bibitem{cekic2021robust} M. Cekic, S. Gopalakrishnan, and U. Madhow, ``Wireless fingerprinting via deep learning: The impact of confounding factors,'' in \emph{Proc. Asilomar Conf. Signals, Syst., Comput.}, 2021, pp. 677--684.
\bibitem{zhang2023radio} J. Zhang, R. Woods, M. Sandell, M. Valkama, A. Marshall, and J. Cavallaro, ``Radio frequency fingerprint identification for narrowband systems, modelling and classification,'' \emph{IEEE Trans. Inf. Forensics Security}, vol. 16, pp. 3974--3987, Jul. 2021.
\bibitem{xie2024towards} L. Xie, L. Peng, and J. Zhang, ``Towards robust RF fingerprint identification using spectral regrowth and carrier frequency offset,'' in \emph{Proc. IEEE INFOCOM}, 2025, pp. 1--10.
\bibitem{ding2018specific} L. Ding, S. Wang, F. Wang, and W. Zhang, ``Specific emitter identification via convolutional neural networks,'' \emph{IEEE Commun. Lett.}, vol. 22, no. 12, pp. 2591--2594, Dec. 2018.
\bibitem{wan2025sigmix} H. Wan, F. Shi, Y. Wang, Q. Xuan, Y. Lin, and G. Gui, ``SigMix: Robust specific emitter identification method enhanced by cross-time and cross-receiver mixing augmentation,'' \emph{IEEE Internet Things J.}, vol. 12, no. 12, pp. 21339--21349, Jun. 2025.
\bibitem{ahmed2014blind} A. Ahmed, B. Recht, and J. Romberg, ``Blind deconvolution using convex programming,'' \emph{IEEE Trans. Inf. Theory}, vol. 60, no. 3, pp. 1711--1732, Mar. 2014.
\bibitem{ling2015self} S. Ling and T. Strohmer, ``Self-calibration and biconvex compressive sensing,'' \emph{Inverse Problems}, vol. 31, no. 11, Art. no. 115002, Nov. 2015.
\bibitem{ling2017blind} S. Ling and T. Strohmer, ``Blind deconvolution meets blind demixing: Algorithms and performance bounds,'' \emph{IEEE Trans. Inf. Theory}, vol. 63, no. 7, pp. 4497--4520, Jul. 2017.
\bibitem{li2019rapid} X. Li, S. Ling, T. Strohmer, and K. Wei, ``Rapid, robust, and reliable blind deconvolution via nonconvex optimization,'' \emph{Appl. Comput. Harmon. Anal.}, vol. 47, no. 3, pp. 893--934, Nov. 2019.
\bibitem{chi2019nonconvex} Y. Chi, Y. M. Lu, and Y. Chen, ``Nonconvex optimization meets low-rank matrix factorization: An overview,'' \emph{IEEE Trans. Signal Process.}, vol. 67, no. 20, pp. 5239--5269, Oct. 2019.
\bibitem{gregor2010learning} K. Gregor and Y. LeCun, ``Learning fast approximations of sparse coding,'' in \emph{Proc. ICML}, 2010, pp. 399--406.
\bibitem{chen2018theoretical} X. Chen, J. Liu, Z. Wang, and W. Yin, ``Theoretical linear convergence of unfolded ISTA and its practical weights and thresholds,'' in \emph{Proc. NeurIPS}, 2018, pp. 9061--9071.
\bibitem{liu2019alista} J. Liu, X. Chen, Z. Wang, and W. Yin, ``ALISTA: Analytic weights are as good as learned weights in LISTA,'' in \emph{Proc. ICLR}, 2019.
\bibitem{samuel2019learning} N. Samuel, T. Diskin, and A. Wiesel, ``Learning to detect,'' \emph{IEEE Trans. Signal Process.}, vol. 67, no. 10, pp. 2554--2564, May 2019.
\bibitem{he2020model} H. He, C.-K. Wen, S. Jin, and G. Y. Li, ``Model-driven deep learning for MIMO detection,'' \emph{IEEE Trans. Signal Process.}, vol. 68, pp. 1702--1715, Feb. 2020.
\bibitem{borgerding2017amp} M. Borgerding, P. Schniter, and S. Rangan, ``AMP-inspired deep networks for sparse linear inverse problems,'' \emph{IEEE Trans. Signal Process.}, vol. 65, no. 16, pp. 4293--4308, Aug. 2017.
\bibitem{yang2016deep} Y. Yang, J. Sun, H. Li, and Z. Xu, ``Deep ADMM-Net for compressive sensing MRI,'' in \emph{Proc. NeurIPS}, 2016, pp. 10--18.
\bibitem{balatsoukas2019deep} A. Balatsoukas-Stimming and C. Studer, ``Deep unfolding for communications systems: A survey and some new directions,'' in \emph{Proc. IEEE SiPS}, 2019, pp. 266--271.
\bibitem{monga2021algorithm} V. Monga, Y. Li, and Y. C. Eldar, ``Algorithm unrolling: Interpretable, efficient deep learning for signal and image processing,'' \emph{IEEE Signal Process. Mag.}, vol. 38, no. 2, pp. 18--44, Mar. 2021.
\bibitem{shlezinger2023model} N. Shlezinger, J. Whang, Y. C. Eldar, and A. G. Dimakis, ``Model-based deep learning,'' \emph{Proc. IEEE}, vol. 111, no. 5, pp. 465--499, May 2023.
\bibitem{mallat2012group} S. Mallat, ``Group invariant scattering,'' \emph{Commun. Pure Appl. Math.}, vol. 65, no. 10, pp. 1331--1398, Oct. 2012.
\bibitem{bruna2013invariant} J. Bruna and S. Mallat, ``Invariant scattering convolution networks,'' \emph{IEEE Trans. Pattern Anal. Mach. Intell.}, vol. 35, no. 8, pp. 1872--1886, Aug. 2013.
\bibitem{kakarala2012bispectrum} R. Kakarala, ``The bispectrum as a source of phase-sensitive invariants for Fourier descriptors: A group-theoretic approach,'' \emph{J. Math. Imaging Vis.}, vol. 44, no. 3, pp. 341--353, Nov. 2012.
\bibitem{bandeira2017estimation} A. S. Bandeira, B. Blum-Smith, J. Kileel, A. Perry, J. Weed, and A. S. Wein, ``Estimation under group actions: Recovering orbits from invariants,'' \emph{Appl. Comput. Harmon. Anal.}, vol. 66, pp. 236--319, Sep. 2023.
\bibitem{bendory2023compact} T. Bendory, D. Edidin, W. Leeb, and N. Sharon, ``Dihedral multi-reference alignment,'' \emph{IEEE Trans. Inf. Theory}, vol. 68, no. 5, pp. 3489--3499, May 2022.
\bibitem{schreier2010statistical} P. J. Schreier and L. L. Scharf, \emph{Statistical Signal Processing of Complex-Valued Data: The Theory of Improper and Noncircular Signals}. Cambridge, U.K.: Cambridge Univ. Press, 2010.
\bibitem{vandenbos1994complex} A. van den Bos, ``Complex gradient and Hessian,'' \emph{IEE Proc. Vis. Image Signal Process.}, vol. 141, no. 6, pp. 380--382, Dec. 1994.
\bibitem{kay1993fundamentals} S. M. Kay, \emph{Fundamentals of Statistical Signal Processing: Estimation Theory}. Upper Saddle River, NJ, USA: Prentice-Hall, 1993.
\end{thebibliography}
\end{document}